\documentclass[journal]{IEEEtran}
\ifCLASSINFOpdf
\else
\fi

\usepackage[utf8]{inputenc}

\usepackage{pifont}
\usepackage{amsmath}

\usepackage{amssymb}
\def\x{{\mathbf x}}
\def\L{{\cal L}}
\usepackage{booktabs}
\usepackage{multirow}
\usepackage{multicol}
\def\x{{\mathbf x}}
\def\L{{\cal L}}
\usepackage{graphicx}
\usepackage{color}
\usepackage{float}
\restylefloat{table}
\usepackage{placeins}
\usepackage{epstopdf}

\newcommand\supinf{\underset{H_0}{\overset{H_1}{\gtrless}}}

\usepackage{hyperref}
\DeclareMathOperator{\Tr}{Tr}

\usepackage[none]{hyphenat}

\begin{document}
\title{Sparse and Low-Rank Matrix Decomposition for Automatic Target Detection in Hyperspectral Imagery}

\author{Ahmad W. Bitar,~\IEEEmembership{Student Member,~IEEE,}
        Loong-Fah~Cheong,~\IEEEmembership{Member,~IEEE,}
        and~Jean-Philippe~Ovarlez,~\IEEEmembership{Member,~IEEE}
\thanks{A. W. Bitar is with SONDRA Lab, CentraleSup\'elec, Universit\'e Paris-Saclay, F-91190 Gif-sur-Yvette, France (e-mail: ahmad.bitar@centralesupelec.fr).}
\thanks{L.-F. Cheong is with the Electrical and Computer Engineering Department, National University of Singapore, Singapore 119077 (e-mail: eleclf@nus.edu.sg).}
\thanks{J.-P. Ovarlez is with ONERA, Universit\'e Paris-Saclay (DEMR/MATS), 91120 Palaiseau, France, and also with the SONDRA Lab, CentraleSup\'elec, Universit\'e Paris-Saclay, F-91190 Gif-sur-Yvette, France (e-mail: jeanphilippe.ovarlez@centralesupelec.fr).}
}

\maketitle

\begin{abstract} 
Given a target prior information, our goal is to propose a method for automatically separating targets of interests from the background in hyperspectral imagery. More precisely, we regard the given hyperspectral image (HSI) as being made up of the sum of low-rank background HSI and a sparse target HSI that contains the targets based on a pre-learned target dictionary constructed from some online spectral libraries. Based on the proposed method, two strategies are briefly outlined and evaluated to realize the target detection on both synthetic and real experiments. 
\end{abstract}


\IEEEpeerreviewmaketitle

\section{Introduction}
\label{sec:intro}

An airborne hyperspectral imaging sensor is capable of simultaneously acquiring the same spatial scene in a contiguous and multiple narrow (0.01 - 0.02 $\mu$m) spectral wavelength (color) bands \cite{Shaw02, manolakis2003hyperspectral, manolakis_lockwood_cooley_2016, 7564440, ZHANG20153102}. When all the spectral bands are stacked together, the result is a hyperspectral image (HSI) whose crosssection is a function of the spatial coordinates and its depth is a function of wavelength. Hence, an HSI is a 3-D data cube having two spatial dimensions and one spectral dimension. Each band of the HSI corresponds to an image of the surface covered by the field of view of the hyperspectral sensor; whereas each ``pixel'' in the HSI is a $p$-dimensional vector, $\mathbf{x}\in\mathbb{R}^p$ ($p$ stands for the total number of spectral bands), consisting of a spectrum characterizing the materials within the pixel. 
The HSI usually contains both pure and mixed pixels. A pure pixel contains only one single material, whereas a mixed pixel contains multiple materials, with its spectral signature representing the aggregate of all the materials in the corresponding spatial location. The latter situation often arises because HSIs are collected hundreds to thousands of meters away from an object so that the object becomes smaller than the size of a pixel. Other scenarios might involve, for example, a military target hidden under foliage or covered with camouflage material.

With the rich information afforded by the high spectral dimensionality, hyperspectral imagery has found many applications in various fields, such as agriculture \cite{Patel2001, Datt2003}, mineralogy \cite{Lehmann2001}, military \cite{manolakis2002detection, Stein02, 4939406}, and, in particular, target detection \cite{Shaw02, manolakis2003hyperspectral, Manolakis14, Manolakis09, manolakis2002detection, 7739987, 7165577, 8069001}. Usually, the detection is built using a binary hypothesis test that chooses between the following competing null and alternative hypothesis: target absent ($H_0$), that is, the test pixel $\mathbf{x}$ consists only of background, and target present ($H_1$), where $\mathbf{x}$ may be either fully or partially occupied by the target material. 
It is well known that the signal model for hyperspectral test pixels is fundamentally different from the additive model used in radar and communications applications \cite{Manolakis09, manolakis_lockwood_cooley_2016}. We can regard each test pixel $\mathbf{x}$ as being made up of $\mathbf{x} = \alpha \, \mathbf{t}$ + $(1-\alpha) \, \mathbf{b}$, where $0 \leq \alpha \leq 1$ designates the target fill-fraction, $\mathbf{t}$ is the spectrum of the target, and $\mathbf{b}$ is the spectrum of the background. When $\alpha = 1$, the pixel $\mathbf{x}$ is fully occupied by the target material and is usually referred to as the full or resolved target pixel. When $0<\alpha<1$, the pixel $\mathbf{x}$ is partially occupied by the target material and is usually referred to as the subpixel or unresolved target.

Different Gaussian-based target detectors (e.g. Matched Filter \cite{Manolakis00, Nasrabadi08}, Normalized Matched Filter \cite{kraut1999cfar}, and Kelly detector \cite{Kelly86}) have been developed. In these classical detectors, the target of interest to detect is known, that is, its spectral signature is fully provided to the user. However, these detectors present several limitations in real-world hyperspectral imagery.
First, they depend on the unknown covariance matrix (of the background surrounding the test pixel) whose entries have to be carefully estimated, especially in large dimensions \cite{LEDOIT2004365, LedoitHoney, AhmadCamsap2017}, and to ensure success under different environments \cite{5606730, 6884641, 6894189}.
Second, there is always an explicit assumption (specifically, Gaussian) on the statistical distribution characteristics of the observed data. For instance, most materials are treated as Lambertian because their bidirectional reflectance distribution function characterizations are usually not available, but the actual reflection is likely to have both a diffuse and a specular component. This latter component would result in gross corruption of the data. In addition, spectra from multiple materials are usually assumed to interact according to a linear mixing model; nonlinear mixing effects are not represented and will contribute to another source of noise.
Finally, the use of only a single reference spectrum for the target of interest may be inadequate since in real-world hyperspectral imagery, various effects that produce variability to the material spectra (e.g. atmospheric conditions, sensor noise, and material composition) are inevitable. For instance, target signatures are typically measured in laboratories or in the field with handheld spectrometers that are at most a few inches from the target surface. HSIs, however, are collected at huge distances away from the target and have significant atmospheric effects present.

To more effectively separate these non-Gaussian noise from signal and to have a target detector that is invariant to atmospheric effects, dictionaries of target and background have been developed (denoted as $\mathbf{A}_t$ and $\mathbf{A}_b$ in this paper), and the test signal is then modeled as a sparse linear combination of the prototype signals taken from the dictionaries \cite{chen11, Chen11b, Zhang15}. This sparse representation approach can alleviate the spectral variability caused by atmospheric effects and can also better deal with a greater range of noise phenomena. This paper falls under this broad family of dictionary-based approach. 
\\
Although these dictionary-based-methods can, in principle, address all the aforementioned limitations, the main drawback is that they usually lack a sufficiently universal dictionary, especially for the background $\mathbf{A}_b$; some form of in-scene adaptation would be desirable. Chen {\it et al.} \cite{chen11, Chen11b} have demonstrated in their sparse representation approach that using an adaptive scheme (a local method) to construct $\mathbf{A}_b$ usually yields better target detection results than with a global dictionary generally constructed from some background materials (e.g. trees, grass, road, buildings, and vegetation). This is to be expected since the subspace spanned by the background dictionary $\mathbf{A}_b$ becomes adaptive to the local statistics.
Zhang {\it et al.} \cite{Zhang15} have used the same adaptive scheme in their sparse representation-based binary hypothesis (SRBBH) approach. 

\begin{figure}[!tbp]
\centering
\begin{minipage}[b]{0.39\textwidth}
\includegraphics[width=\textwidth]{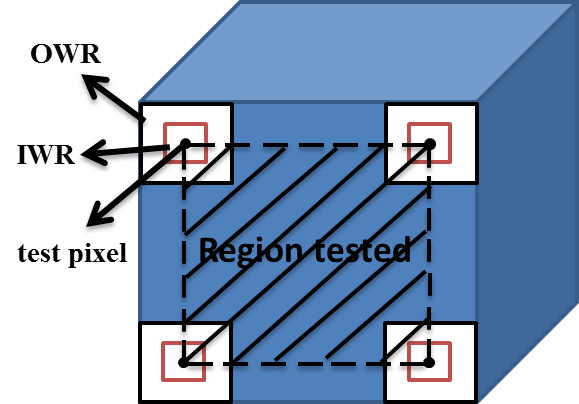}
\caption{Sliding dual concentric window across the HSI.}
\label{fig:dual_window}
\end{minipage}
\end{figure}

In fact, the construction of a locally adaptive dictionary $\mathbf{A}_b$ is a very challenging problem since a contamination of it by the target pixels can potentially affect the target detection performance. Usually, the adaptive scheme is based on a dual concentric window centered on the test pixel (see Fig. \ref{fig:dual_window}), with an inner window region (IWR) centered within an outer window region (OWR), and only the pixels in the OWR will constitute the samples for $\mathbf{A}_b$. In other words, if the size of OWR is $m \times m$ and the size of IWR is $l \times l$, where $l < m$, then the total number of pixels in the OWR that will form $\mathbf{A}_b$ is $m^2 - l^2$. Clearly, the dimension of IWR is very important and has a strong impact on the target detection performance since it aims to enclose the targets of interests to be detected. It should be set larger than or equal to the size of all the desired targets of interests in the corresponding HSI, so as to exclude the target pixels from erroneously appearing in $\mathbf{A}_b$. However, information about the target size in the image is usually not at our disposal. It is also very unwieldy to set this size parameter when the target could be of an irregular shape (e.g., searching for lost plane parts of a missing aircraft). Another tricky situation is when there are multiple targets in close proximity in the image (e.g., military vehicles in long convoy formation). 

In this paper, we handle the aforementioned challenges in constructing $\mathbf{A}_b$ by providing a method capable of removing the targets from the background, and hence, avoiding the use of an IWR to construct $\mathbf{A}_b$ as well as dealing with a larger range of target size, shape, number, and placement in the image. Based on a modification of the recently developed robust principal component analysis (RPCA) \cite{Candes11}, our method decomposes an input HSI into a background HSI (denoted by $\mathbf{L}$) and a sparse target HSI (denoted by $\mathbf{E}$) that contains the targets of interests. 
\\
While we do not need to make assumptions about the size, shape, or number of the targets, our method is subject to certain generic constraints that make less specific assumption on the background or the target. These constraints are similar to those used in RPCA \cite{Candes11, NIPS2009_3704}, including: 1) the background is not too heavily cluttered with many different materials with multiple spectra so that the background signals should span a low-dimensional subspace, a property that can be expressed as the low-rank condition of a suitably formulated matrix \cite{ChenYu, Zhang15b, 7322257, 8260545, 8126244, Ahmad2017a}; 
2) the total image area of all the target(s) should be small relative to the whole image (i.e., spatially sparse), e.g., several hundred pixels in a million pixel image, though there is no restriction on a target shape or the proximity between the targets.

\begin{figure*}[!tbp]
\minipage{0.2\textwidth}
\includegraphics[width=\linewidth]{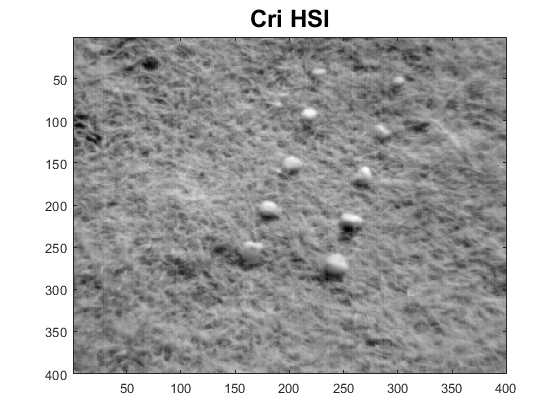}
\endminipage\hfill
\centering
\minipage{0.2\textwidth}
\includegraphics[width=\linewidth]{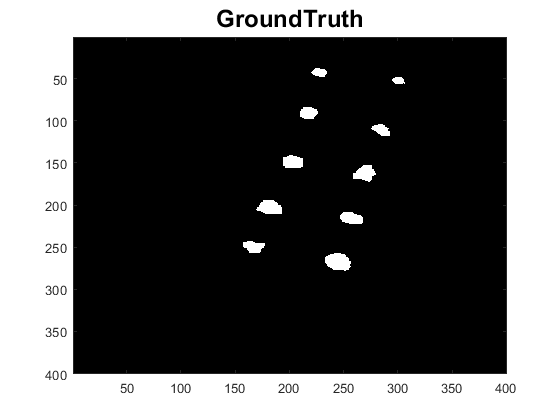}
\endminipage\hfill
\minipage{0.2\textwidth}
\includegraphics[width=\linewidth]{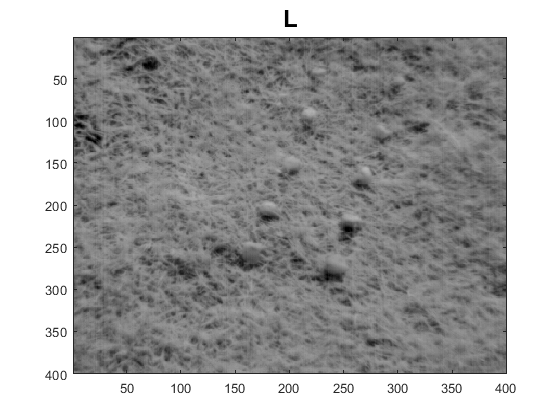}
\endminipage\hfill
\minipage{0.2\textwidth}
\includegraphics[width=\linewidth]{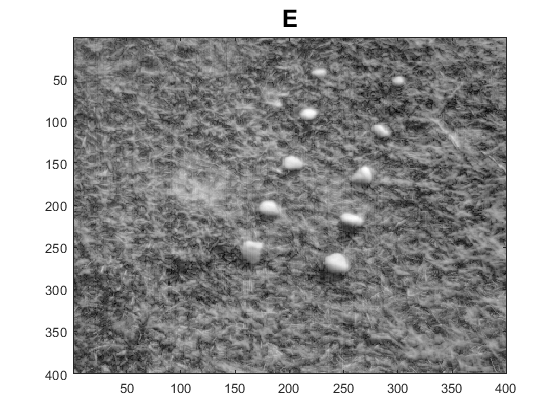}
\endminipage\hfill
\minipage{0.2\textwidth}
\includegraphics[width=\linewidth]{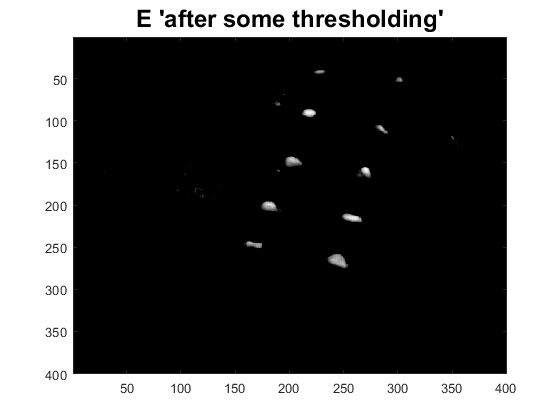}
\endminipage\hfill
\centering
\minipage{0.2\textwidth}
\includegraphics[width=\linewidth]{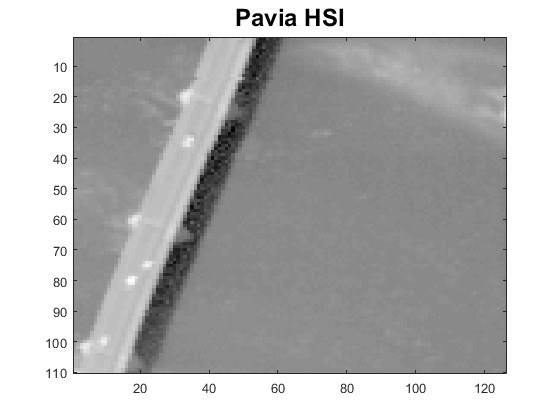}
\endminipage\hfill
\centering
\minipage{0.2\textwidth}
\includegraphics[width=\linewidth]{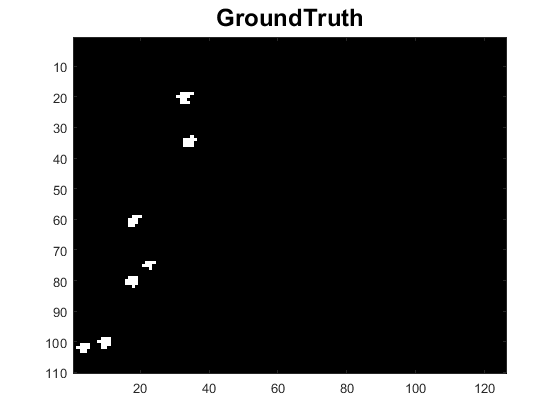}
\endminipage\hfill
\minipage{0.2\textwidth}
\includegraphics[width=\linewidth]{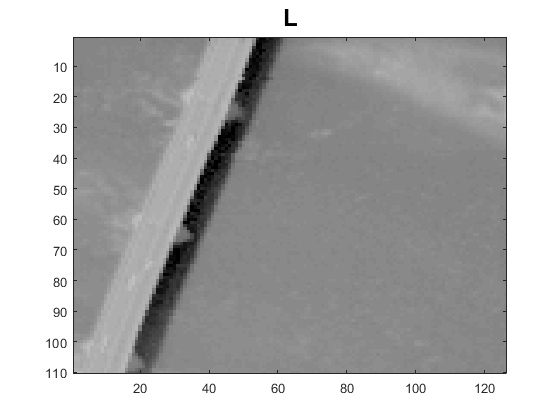}
\endminipage\hfill
\minipage{0.2\textwidth}
\includegraphics[width=\linewidth]{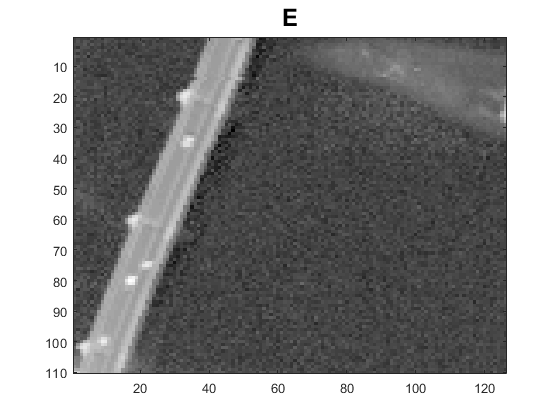}
\endminipage\hfill
\minipage{0.2\textwidth}
\includegraphics[width=\linewidth]{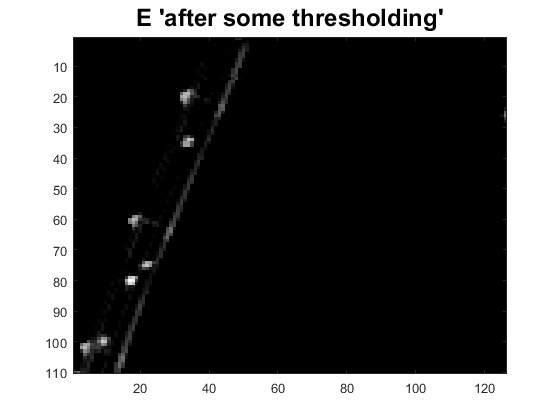}
\endminipage\hfill
\centering
\minipage{0.2\textwidth}
\includegraphics[width=\linewidth]{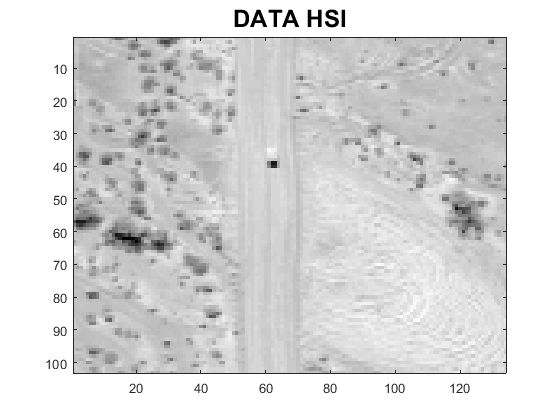}
\endminipage\hfill
\minipage{0.2\textwidth}
\includegraphics[width=\linewidth]{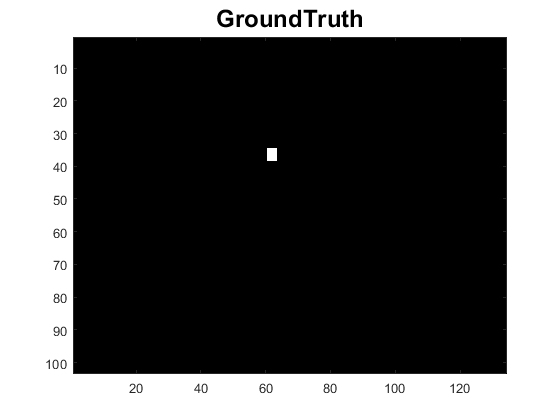}
\endminipage\hfill
\minipage{0.2\textwidth}
\includegraphics[width=\linewidth]{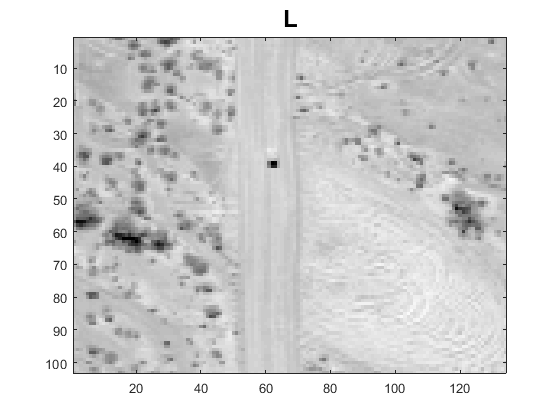}
\endminipage\hfill
\centering
\minipage{0.2\textwidth}
\includegraphics[width=\linewidth]{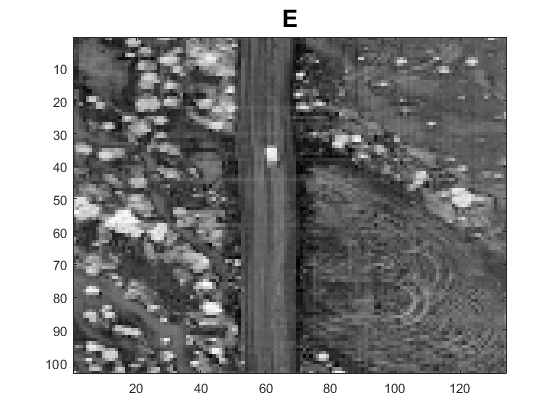}
\endminipage\hfill
\minipage{0.2\textwidth}
\includegraphics[width=\linewidth]{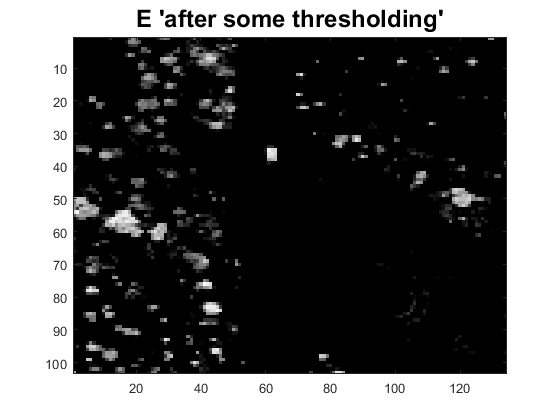}
\endminipage\hfill
\minipage{0.2\textwidth}
\includegraphics[width=\linewidth]{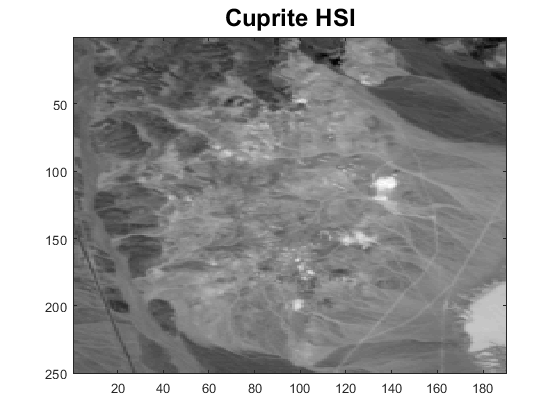}
\endminipage\hfill
\minipage{0.2\textwidth}
\includegraphics[width=\linewidth]{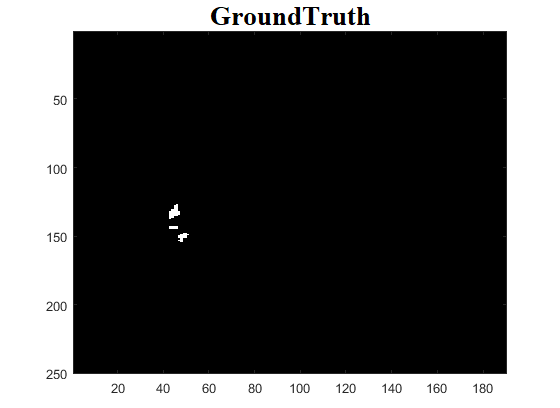}
\endminipage\hfill
\minipage{0.2\textwidth}
\includegraphics[width=\linewidth]{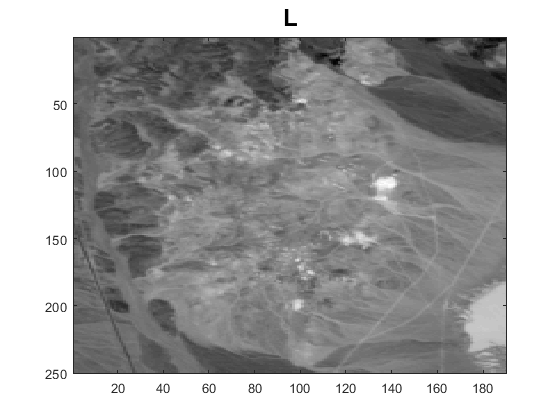}
\endminipage\hfill
\minipage{0.2\textwidth}
\includegraphics[width=\linewidth]{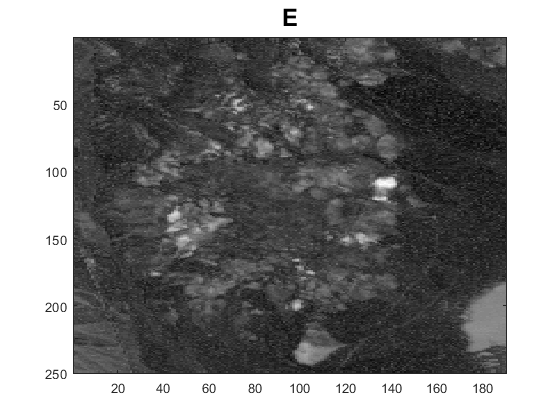}
\endminipage\hfill
\minipage{0.2\textwidth}
\includegraphics[width=\linewidth]{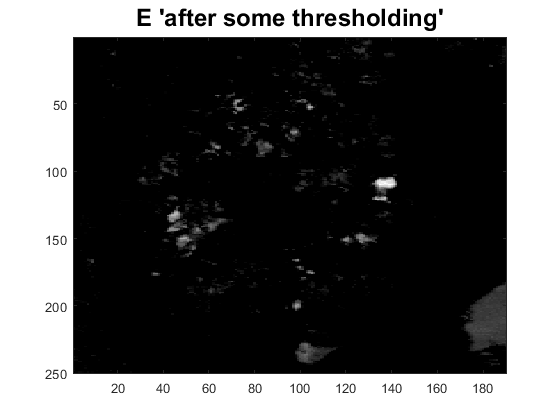}
\endminipage
\caption{Columns from left to right: original HSI (mean power in db), the GroundTruth image for the target of interest, the low-rank background HSI $\mathbf{L}$ (mean power in db), the sparse target HSI $\mathbf{E}$ (mean power in db), and the sparse target HSI $\mathbf{E}$ after some thresholding (mean power in db).
}
\label{fig:SPCP_example}
\end{figure*}

Our method also assumes that the target spectra are available to the user and that the atmospheric influence can be accounted for by the target dictionary $\mathbf{A}_t$. This prelearned target dictionary $\mathbf{A}_t$ is used to cast the general RPCA into a more specific form, specifically, we further factorize the sparse component $\mathbf{E}$ from RPCA into the product of $\mathbf{A}_t$ and a sparse activation matrix $\mathbf{C}$ \cite{8462257}. This modification is essential to disambiguate the true targets from other small objects, as the following discussion will show. In our application, there are often other small, heterogeneous, high contrast regions that are nontargets. These would have been deemed as outliers (targets) under the general RPCA framework. Compounding the decomposition is also the often uniform material present in most targets, which means that they would contribute only a small increase in the rank of the background HSI $\mathbf{L}$ if they were to be grouped in the background HSI. Indeed, some other heterogeneous nontarget objects or specular highlights may contribute a larger increase in rank, and thus, they are more liable to be treated as outliers (targets) in the decomposition under the general RPCA. In other words, there is a substantial overlap between the $\mathbf{L}$ and $\mathbf{E}$ for the general RPCA to be well posed or work well.

Let us take an example in Fig. \ref{fig:SPCP_example} that uses the RPCA model solved via stable principal component pursuit \cite{SPCPCandes} on four real HSIs. As can be seen, despite the effort to individually tune the parameters for the best separation for each of the four images, it is not possible to obtain a clean separation. And even with a lot of false alarms in the sparse target image, the background is still not completely cleansed of the target. 
\begin{enumerate}
\item The first HSI \cite{5346879, Zhang15b} (Cri HSI) is acquired by the Nuance Cri hyperspectral sensor. It covers an area of $400 \times 400$ pixels with 46 spectral bands in wavelengths ranging from 650 to 1100 nm. It contains ten rocks targets in a simple background and thus poses no problem for the general RPCA. However, the other images represent a more complex background.
\item The second HSI (Pavia HSI) is a selected small zone from Pavia Center City. It is a {\color{red}$110 \times 126$} image and consists of 102 spectral bands in wavelengths ranging from 430 to 860 nm. The main background materials are bridge and water. There are some vehicles
on the bridge and the bare soil near the bridge pier, and hence, they are considered as targets to be detected. We can obviously observe that both the vehicles on the bare soil and the bridge pier are being deposited in the sparse image.
\item The third HSI \cite {Terreaux15} (Data HSI) which is a {\color{red}$103 \times 134$} image and consisting of 167 spectral bands, depicts a scrubby terrain with small heterogeneous regions comprised of trees and one vehicle, the latter is being the target of interest in this case. We observe that both the vehicle and trees are being deposited in the sparse target image.
\item The fourth HSI \cite{Swayze10245, JGRE:JGRE1642} (Cuprite HSI) is a region of the Cuprite mining district area, of size {\color{red}$250 \times 190$} pixels and consisting of 186 spectral bands in wavelengths ranging from 0.4046 to 2.4573 $\mu$ m. In this small zone area, three buddingtonite outcrops (spectrally dominated by buddingtonite) are considered as targets, and their locations are shown in the GroundTruth\footnote{Note that there may also be smaller buddingtonite outcrops in the northeastern (NE) quadrant of the eastern alteration center, but they are spectrally dominated by alunite's absorption.}. It has been noted by Swayze et. al. in \cite{Swayze10245} that the ammonia in buddingtonite has a distinct N-H combination absorption at 2.12 $\mu$m, a position similar to that of the cellulose absorption in dried vegetation, from which it can be distinguished based on its narrower bandwidth and asymmetry. Hence, the buddingtonite 2.12 $\mu$m combination absorption is unique in wavelength location relative to those of most other minerals in the image (that is, it is easily recognized based on its unique 2.12 $\mu$m absorption band). This might be a reason of why the general RPCA is able to find those buddingtonite outcrops in addition to the small heterogeneous and high contrast regions, which are also deposited in the sparse target image. 
In addition, one can imagine that the lignin N-H absorption in vegetation would look somewhat like the N-H combination absorption in buddingtonite, but that using more spectral bands better differentiates buddingtonite from lignin in plants. Both are relatively broad absorptions isolated in wavelength space from other absorptions. However, there is actually vegetation at Cuprite, probably from 10 to 15\% ground cover, though the buddingtonite areas in this image zone are relatively vegetation-free. Thus, the three buddingtonite outcrops in this image zone can be considered to be homogeneously surrounded by areas with more vegetation on the western side of the eastern alteration center.
\\
Note that in the experiments later, only the fourth HSI is evaluated since for the first three HSIs, there are no available samples for the targets in the online spectral libraries.
\end{enumerate}
In this regard, the incorporation of the target dictionary prior can, we feel, greatly help in identifying the true targets and separate them from the background. From the proposed model, we aim to use the background HSI $\mathbf{L}$ for a more accurate construction of $\mathbf{A}_b$, following which various dictionary-based-methods can be used to carry out a more elaborate binary hypothesis test. Via the background HSI $\mathbf{L}$, a locally adaptive dictionary $\mathbf{A}_b$ can be constructed without the need of using an IWR and also avoiding contamination by the target pixels.
\\
An alternative strategy would be to directly use the target HSI (the product of $\mathbf{A}_t$ and the sparse activation matrix $\mathbf{C}$) as a detector. That is, we detect the nonzero entries of the sparse target image, and the targets are deemed to be present at these nonzero supports. 

This paper is structured along the following lines. First comes an overview of some related works in section \ref{sec:RelatedWorks}. In Section \ref{sec:Maincontribution}, the proposed decomposition model and the two strategies that realize the target detection are briefly outlined. Section \ref{sec:experimentss} presents both synthetic and real experiments to gauge the effectiveness of the two outlined strategies. This paper ends with a summary of the work and some directions for future work. 

{\em Summary of Main Notations:} Throughout this paper, we depict vectors in lowercase boldface letters and matrices in uppercase boldface letters. The notation $(.)^T$ and $\mathrm{Tr}(.)$ stand for the transpose and trace of a matrix, respectively. In addition, $\mathrm{rank}(.)$ is for the rank of a matrix. A variety of norms on matrices will be used. For instance, $\mathbf{M}$ is a matrix, $[\mathbf{M}]_{:,j}$ is the $j$th column. The matrix $l_{2,0}$, $l_{2,1}$ norms are defined by $\left\Vert\mathbf{M}\right\Vert_{2,0} = \# \left\{ j \, : \, \left\Vert\left[\mathbf{M}\right]_{:,j}\right\Vert_2 \, \not= \, 0\right\}$, and $\left\Vert\mathbf{M}\right\Vert_{2,1} = \displaystyle\sum_j \left\Vert\left[\mathbf{M}\right]_{:,j}\right\Vert_2$, respectively. The Frobenius norm and the nuclear norm (the sum of singular values of a matrix) are denoted by $\left\Vert\mathbf{M}\right\Vert_F$ and $\left\Vert\mathbf{M}\right\Vert_* = \Tr\left(\mathbf{M}^T \,\mathbf{M}\right)^{(1/2)}$, respectively.


\section{Related works}
\label{sec:RelatedWorks}

Besides the generic RPCA and its variants discussed in Section \ref{sec:intro}, there have been other modifications of RPCA.
For example, the generalized model of RPCA, named the low-rank representation (LRR) \cite{6180173}, allows the use of a subspace basis as a dictionary or just uses self-representation to obtain the LRR. The major drawback in LRR is that the incorporated dictionary has to be constructed from the background and to be pure from the target samples. This challenge is similar to our background dictionary $\mathbf{A}_b$  construction problem. If we use the self-representation form of LRR, the presence of a target in the input image may only bring about a small increase in rank (as discussed in Section \ref{sec:intro}) and thus be retained in the background.

In the earliest models using a low-rank matrix to represent background \cite{Candes11, NIPS2009_3704, SPCPCandes}, no prior knowledge on the target was considered. In some applications, such as speech enhancement and hyperspectral imagery, we may expect some prior information about the target of interest, which can be provided to the user. Thus, incorporating this information about the target into the separation scheme in the general RPCA model should allow us to potentially improve the target extraction performance. For example, Chen and Ellis \cite{6701883} and Sun and Qin \cite{7740039} proposed a speech enhancement system by exploiting the knowledge about the likely form of the targeted speech. This was accomplished by factorizing the sparse component from RPCA into the product of a dictionary of target speech templates and a sparse activation matrix. The proposed methods in \cite{6701883} and \cite{7740039} typically differ on how the fixed target dictionary of speech spectral templates is constructed.
Our model separation in section \ref{sec:Maincontribution} is very related to \cite{6701883} and \cite{7740039}. In real-world hyperspectral imagery, the prior target information may not be only related to its spatial properties (e.g.  size, shape, and texture), which is usually not at our disposal, but also to its spectral shape signature. The latter usually hinges on the nature of the given HSI where the spectra of the targets of interests present have been already measured by some laboratories or with some handheld spectrometers.

In addition, by using physical models and the MODTRAN atmospheric modeling program \cite{Berk89}, a number of samples for a specific target can be generated under various atmospheric conditions.

Now, we provide an overview of the SRBBH detector that will be used for evaluation throughout the experiments later.
\subsection{Overview of the SRBBH Detector \cite{Zhang15}}
The SRBBH detector is defined as follows:
\begin{equation}
\label{SRBBH}
D_{SRBBH}(\mathbf{x}) = ||\mathbf{x} - \mathbf{A}_b \,\hat{\boldsymbol{\theta}}||_2 - \left\Vert\mathbf{x} - \mathbf{A} \, \hat{\boldsymbol{\gamma}}\right\Vert_2 \,,
\end{equation}
with
\begin{eqnarray*}
\footnotesize
&\hat{\boldsymbol{\theta}} = \underset{\boldsymbol{\theta}} {\mathrm{argmin}} \left\Vert\mathbf{x} - \mathbf{A}_b \, \boldsymbol{\theta}\right\Vert_2 ~~ \text{s.t.}~~ \left\Vert\boldsymbol{\theta}\right\Vert_0 \leq k_0 \,,\\
&\hat{\boldsymbol{\gamma}} = \underset{\boldsymbol{\gamma}} {\mathrm{argmin}} \left\Vert\mathbf{x} - \mathbf{A} \, \boldsymbol{\gamma}\right\Vert_2 ~~ \text{s.t.}~~ \left\Vert\boldsymbol{\gamma}\right\Vert_0 \leq k'_0 \,.
\end{eqnarray*}
where $\mathbf{A}_b \in \mathbb{R}^{p \times N_b}$, $\mathbf{A} = \left[\mathbf{A}_b \, \mathbf{A}_t\right] \in \mathbb{R}^{p \times (N_b + N_t)}$. Both $\boldsymbol{\theta} \in \mathbb{R}^{N_b}$ and $\boldsymbol{\gamma} \in \mathbb{R}^{N_b + N_t}$ tend to be a sparse vectors. Actually, $k_0$ and $k'_0$ are a given upper bound on the sparsity level \cite{Tropp10}. For simplicity, and as in \cite{Zhang15}, $k_0$ and $k'_0$ are set equally to each other. In the experiments later, we set $k_0=k'_0 = 8$. 
\\
In this paper, we solve each of $\boldsymbol{\theta}$ and $\boldsymbol{\gamma}$ using the orthogonal matching pursuit \cite{Tropp06} greedy algorithm. If $D_{SRBBH}(\mathbf{x}) > \eta$ with $\eta$ being a prescribed threshold value, then $\mathbf{x}$ is declared as a target; otherwise, $\mathbf{x}$ will be labeled as background.
\\
In fact, the SRBBH detector was developed very recently to combine the idea of binary hypothesis and sparse representation, obtaining a more complete and realistic model than \cite{chen11}. More precisely, if the test pixel belongs to $H_0$, it will be modeled by the background dictionary $\mathbf{A}_b$ only; otherwise, it will be modeled by the union of $\mathbf{A}_b$ and $\mathbf{A}_t$. This, in fact, yields a competition between the two hypotheses corresponding to the different pixel class label.  


   
\section{Main contribution}
\label{sec:Maincontribution}

\subsection{Problem Formulation} 
Suppose an HSI of size $h \times w \times p$, where $h$ and $w$ are the height and width of the image scene, respectively, and $p$ is the number of spectral bands. 
Let us consider that the given HSI contains $q$ target pixels, $\left\{ \mathbf{x_i}\right\}_{i\in[1,\,q]}$, $\mathbf{x}_i = \alpha_i \, \mathbf{t}_i + (1 - \alpha_i) \, \mathbf{b}_i$ with $0<\alpha_i \leq 1$, where $\mathbf{t}_i$ represents the known target that replaces a fraction $\alpha_i$ of the background $\mathbf{b}_i$ (i.e. at the same spatial location). The remaining ($e-q$) pixels in the given HSI, with $e = h \times w$, are thus only background ($\alpha=0$).
By assuming that all $\left\{\mathbf{t}_i\right\}_{i\in[1,\,q]}$ consist of similar materials, they should be represented by a linear combination of $N_t$ common target samples $\left\{\mathbf{a}^t_{j}\right\}_{j \in [1, \, N_t]}$, where $\mathbf{a}_j^t \in\mathbb{R}^p$ (the superscript $t$ is for target), but weighted with different set of coefficients $\left\{\beta_{i,j}\right\}_{j\in[1, N_t]}$. Thus, each of the $q$ targets is represented as
\begin{equation}
\mathbf{x}_i = \alpha_i \sum\limits_{j=1}^{N_t} \left(\beta_{i,j}\, \mathbf{a}^t_{j} \right) + (1 - \alpha_i) \, \mathbf{b}_i \hspace{0.5cm} i \in [1,q]\, . 
\end{equation}
We rearrange the given HSI into a 2-D matrix $\mathbf{D}\in \mathbb{R}^{e \times p}$, with $e = h \times w$ (by lexicographically ordering the columns). This matrix $\mathbf{D}$, can be decomposed into a low-rank matrix $\mathbf{L}_0$ representing the pure background, a sparse matrix capturing any spatially small signals residing in the known target subspace, and a noise matrix $\mathbf{N}_0$. More precisely, the model used is 
\begin{equation}\label{eq:mod2}
\mathbf{D} = \mathbf{L}_0 + \left(\mathbf{A}_t\, \mathbf{C}_0\right)^T + \mathbf{N}_0\,,
\end{equation}
where $\left(\mathbf{A}_t \, \mathbf{C}_0\right)^T$ is the sparse target matrix, ideally with $q$ nonzero rows representing $\left\{ \alpha_i \, \mathbf{t}^T_i \right\}_{i\in[1,q]}$ , with target dictionary $\mathbf{A}_t \in \mathbb{R}^{p \times N_t}$ having columns representing target samples $\{\mathbf{a}^t_{j}\}_{j \in [1, N_t]}$, and coefficient matrix $\mathbf{C}_0\in\mathbb{R}^{N_t \times e}$ that should be a sparse column matrix, again ideally containing $q$ nonzero columns each representing $\alpha_i \, \left[\beta_{i,1}, \, \cdots, \, \beta_{i, N_t}\right]^T$, $i \in [1, q]$. $\mathbf{N}_0$ is assumed to be independent and identically distributed Gaussian noise with zero mean and unknown standard deviation.
\\
After reshaping $\mathbf{L}_0$, $\left(\mathbf{A}_t\, \mathbf{C}_0\right)^T$ and $\mathbf{N}_0$ back to a cube of size $h \times w \times p$, we call these entities the ``low-rank background HSI'', ``sparse target HSI'', and ``noise HSI'', respectively.
\\
In order to recover the low-rank matrix $\mathbf{L}_0$ and the sparse target matrix $\left(\mathbf{A}_t\, \mathbf{C}_0\right)^T$, we consider the following minimization problem:
\begin{equation}
\label{eq:our_minimization}
\underset{\mathbf{L}, \mathbf{C}} {\mathrm{min}} \,  \left\{\tau \, \mathrm{rank}\left(\mathbf{L}\right)+ \lambda \, \left\Vert\mathbf{C}\right\Vert_{2,0} +  \left\Vert\mathbf{D} - \mathbf{L} - \left(\mathbf{A}_t\, \mathbf{C}\right)^T\right\Vert_F^2 \right\}\,,
\end{equation}
where $\tau$ controls the rank of $\mathbf{L}$, and $\lambda$ the sparsity level in $\mathbf{C}$.

\subsection{Recovering Low-rank Background Matrix and Sparse Target Matrix by Convex Optimization}
Problem \eqref{eq:our_minimization} is NP-hard due to the presence of the rank term and the $\left\Vert.\right\Vert_{2,0}$ term. We relax these terms to their convex proxies, specifically, using nuclear norm $\left\Vert\mathbf{L}\right\Vert_*$  as a surrogate for the $\mathrm{rank}(\mathbf{L})$ term, and the $l_{2,1}$ norm for the $l_{2,0}$ norm.
We now need to solve the following convex minimization problem:
\begin{equation}{\label{eq:convex_model}}
\underset{\mathbf{L}, \mathbf{C}} {\mathrm{min}} \, \left\{\tau \,\left\Vert\mathbf{L}\right\Vert_*+ \lambda \,\left\Vert\mathbf{C}\right\Vert_{2,1} +  \left\Vert\mathbf{D} - \mathbf{L} - \left(\mathbf{A}_t \, \mathbf{C}\right)^T\right\Vert_F^2 \right\} \, .
\end{equation}
Problem \eqref{eq:convex_model} is solved via an alternating minimization of two subproblems. Specifically, at each iteration $k$
\begin{subequations}\label{eq:sub}
\footnotesize
\begin{alignat}{2}
\label{eq:sub1}
\mathbf{L}^{(k)} &= \underset{\mathbf{L}} {\mathrm{argmin}} \, \left\{\left\Vert\mathbf{L} - \left(\mathbf{D} - \left(\mathbf{A}_t \, \mathbf{C}^{(k-1)}\right)^T\right)\right\Vert_F^2 + \tau \,\left\Vert\mathbf{L}\right\Vert_* \, \right\}\,, \\
\label{eq:sub2}
\mathbf{C}^{(k)} &= \underset{\mathbf{C}} {\mathrm{argmin}} \, \left\{\left\Vert\left(\mathbf{D} - \mathbf{L}^{(k)}\right)^T -  \mathbf{A}_t \,  \mathbf{C}\right\Vert_F^2 + \lambda \,\left\Vert\mathbf{C}\right\Vert_{2,1} \, \right\}.
\end{alignat}
\end{subequations}
The minimization subproblems \eqref{eq:sub1} and \eqref{eq:sub2} are convex and each can be solved optimally. Equation \eqref{eq:sub1} is solved via the singular value thresholding operator \cite{SVT2010}. Equation \eqref{eq:sub2} refers to the Lasso problem (if we reshape the matrix $\mathbf{C}$ into a vector) that can be solved by various methods, among which we adopt the alternating direction method of multipliers (ADMM) \cite{Boyd:2011:DOS:2185815.2185816, Lintheaugmented, HiSm1}. 
More precisely, we introduce an auxiliary variable $\mathbf{F}$ into subproblem \eqref{eq:sub2} and recast it into the following form:
\begin{equation}
\label{eq:auxiliary}
\footnotesize
\begin{split}
\hspace{-0.26cm}\left(\mathbf{C}^{(k)}, \mathbf{F}^{(k)}\right) = \underset{\mathbf{C}, \, \mathbf{F}} {\mathrm{argmin}} \, \left\{\left\Vert \left(\mathbf{D} - \mathbf{L}^{(k)}\right)^T - \mathbf{A}_t \, \mathbf{C}\right\Vert_F^2 + \lambda \, \left\Vert\mathbf{F}\right\Vert_{2,1}\right\}
\\
s.t.~~ \mathbf{C} = \mathbf{F} \, .
\end{split}
\end{equation}
Problem \eqref{eq:auxiliary} can then be solved as follows (scaled form of ADMM): 
\begin{equation*}
\footnotesize
\mathbf{C}^{(k)} = \underset{\mathbf{C}} {\mathrm{argmin}} \, \left\{\left\Vert \left(\mathbf{D} - \mathbf{L}^{(k)}\right)^T - \mathbf{A}_t \, \mathbf{C}\right\Vert_F^2 \right. 
\end{equation*}
\begin{equation}
 \hspace{1.2cm} +  \left. \frac{\rho^{(k-1)}}{2}\,\left\Vert \mathbf{C} - \mathbf{F}^{(k-1)} + \frac{1}{\rho^{(k-1)}}\mathbf{Z}^{(k-1)}\right\Vert_F^2 \, \right\} \, ,
\end{equation}
\begin{equation}
\footnotesize
\mathbf{F}^{(k)} = \underset{\mathbf{F}} {\mathrm{argmin}} \, \Bigg\{\lambda \, \left\Vert\mathbf{F}\right\Vert_{2,1} + \frac{\rho^{(k-1)}}{2}\,\left\Vert\mathbf{C}^{(k)} - \mathbf{F} + \frac{1}{\rho^{(k-1)}}\mathbf{Z}^{(k-1)}\right\Vert_F^2\Bigg\} \, ,
\end{equation}
\begin{equation}
\footnotesize
\mathbf{Z}^{(k)} =\mathbf{Z}^{(k-1)} + \rho^{(k-1)} \,\left(\mathbf{C}^{(k)} - \mathbf{F}^{(k)}\right) \, ,
\end{equation}
where $\mathbf{Z} \in \mathbb{R}^{N_t \times e}$ is the Lagrangian multiplier matrix and $\rho$ is a positive scalar. We initialize $\mathbf{L}^{(0)} = \mathbf{C}^{(0)} = \mathbf{F}^{(0)} = \mathbf{Z}^{(0)} = \boldsymbol{0}$, $\rho^{(0)} = 10^{-4}$ and update $\rho^{(k)} = 1.1 \, \rho^{(k-1)}$. The criteria for convergence of problem \eqref{eq:auxiliary} are $\left\Vert\mathbf{C}^{(k)} - \mathbf{F}^{(k)}\right\Vert_F^2 \leq 10^{-6}$.
\\
\\
For problem \eqref{eq:convex_model}, we stop the iteration when the following convergence criterion is satisfied:
\\
\\
$
\frac{ \left\Vert\mathbf{L}^{(k)} - \mathbf{L}^{(k-1)} \right\Vert_F}{ \left\Vert\mathbf{D}\right\Vert_F} \leq \epsilon~\text{and} ~
\frac{ \left\Vert\left(\mathbf{A}_t \, \mathbf{C}^{(k)}\right)^T - \left(\mathbf{A}_t \, \mathbf{C}^{(k-1)}\right)^T\right\Vert_F}{ \left\Vert\mathbf{D}\right\Vert_F} \leq \epsilon \, ,
$
\\
\\
where $\epsilon>0$ is a precision tolerance parameter. In the experiments, we set $\epsilon$ = $10^{-4}$.

\begin{figure}[!tbp]
\centering
\minipage{0.4\textwidth}
  \includegraphics[width=\linewidth]{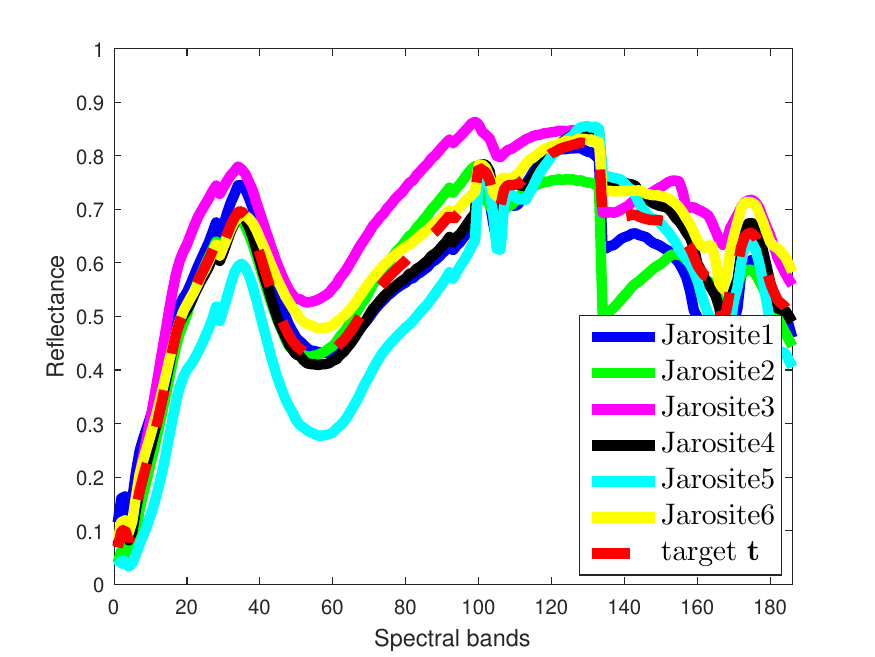}
 \caption{Plot of the six jarosite samples taken from the online USGS spectral library (which will constitute the target dictionary $\mathbf{A}_t$) and the target of interest $\mathbf{t}$ consisting of the mean of the six jarosite samples.}\label{fig:Jarosite_target_samples}
\endminipage
\end{figure}

\subsection{What After the Target and Background Separation} 
\label{sec:what_after}
\vspace{-1.5mm}
Two strategies are available to us to realize the target detection.

1) Strategy One: We use the background HSI $\mathbf{L}$ for a more accurate construction of $\mathbf{A}_b$. For each test pixel in the original HSI, we create a concentric window of size $m \times m$ on the background HSI $\mathbf{L}$, and all the pixels within the window (except the center pixel) will each contribute to one column in $\mathbf{A}_b$. Note that this concentric window amounts to an OWR of size $m \times m$ with IWR of size $1 \times 1$. Next, we make use of the SRBBH detector \cite{Zhang15}\footnote{The reason why we choose the SRBBH detector instead of \cite{chen11} is because it combines the idea of binary hypothesis and sparse representation, obtaining a more complete and realistic model than \cite{chen11}.}, but with the background dictionary $\mathbf{A}_b$ constructed in the preceding manner. Note that for this scheme to work, we do not need a clean separation (by clean separation, we mean that all the targets are present in $\left(\mathbf{A}_t\mathbf{C}\right)^T$ with no false alarms); specifically, we require the entire target fraction to be separated from the background and deposited in the target image, but some of the background objects can also be deposited in the target image. As long as enough signatures of these background objects remain in the background HSI $\mathbf{L}$, $\mathbf{A}_b$ constructed will be adequately representative of the background. 

It is important to mention that the edges of the HSI are not processed and so the images are trimmed in function of the window size. As a result, we will call each of the trimmed images as ``the region tested''. In fact, by taking a large concentric window, a lot of pixels in the image (near the edges) will not be tested. One can imagine how this can become problematic if these excluded pixels from testing contain some or all the targets of interests.
In this regard, after removing the targets from the background by our problem \eqref{eq:convex_model}, a small concentric window will be sufficient to construct an accurate background dictionary $\mathbf{A}_b$, and hence, almost the entire image will be tested.

Note also that we could have constructed $\mathbf{A}_b$ directly from all the pixels in the low-rank background HSI $\mathbf{L}$ (except the pixel that corresponds to the test pixel in the original HSI) without the use of any sliding concentric window. This has the advantage of testing the entire image for the detection (that is, the region tested = the original image).
However, we choose not to do this, as this would result in a substantially larger $\mathbf{A}_b$ size and, therefore, a much increased computational cost. 

2) Strategy Two: We use $\left(\mathbf{A}_t \, \mathbf{C}\right)^T$ directly as a detector. Note that for this scheme to work, we require as few false alarms as possible to be deposited in the target image, but we do not need the target fraction to be entirely removed from the background (that is, a very weak target separation can suffice). As long as enough of the target fractions are moved to the target image such that nonzero support is detected at the corresponding pixel location, it will be adequate for our detection scheme. From this standpoint, we should choose a $\lambda$ that is relatively large so that the target image is really sparse with zero or little false alarms, and only the signals that reside in the target subspace specified by $\mathbf{A}_t$ will be deposited there.

\begin{figure*}[!tbp]
~~~~~~~~~~~~~~~~~~$\mathbf{D}$ ~~~~~~~~~~~~~~~= ~~~~~~~~~~~~~~~$\mathbf{L}$~~~~~~~~~~~~~~~~~  +~~~~~~~~~~~~$\left(\mathbf{A}_t\, \mathbf{C}\right)^T$~~~~~~~~~~~~  +~~~~~~~~~~~~~~~~$\mathbf{N}$
\\
\minipage{0.25\textwidth}
  \includegraphics[width=\linewidth]{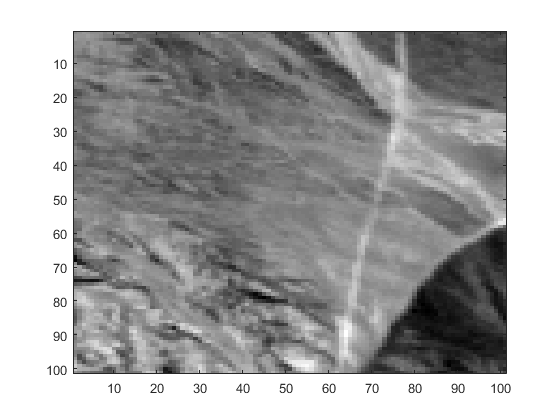}
\endminipage\hfill
\minipage{0.25\textwidth}
  \includegraphics[width=\linewidth]{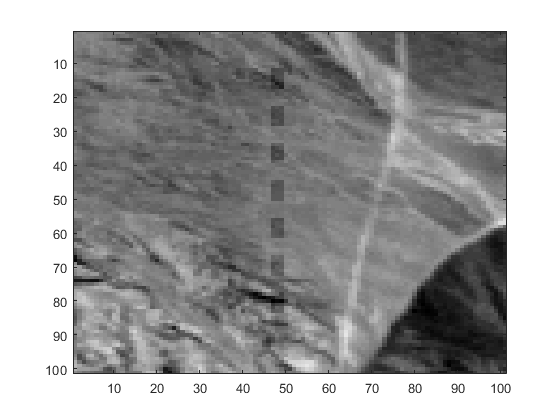}
\endminipage\hfill
\minipage{0.25\textwidth}
  \includegraphics[width=\linewidth]{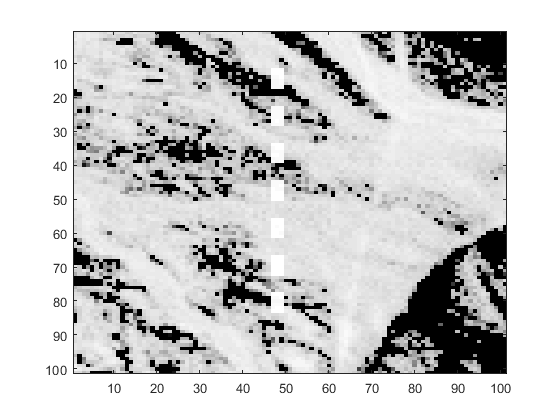}
\endminipage\hfill
\minipage{0.25\textwidth}
  \includegraphics[width=\linewidth]{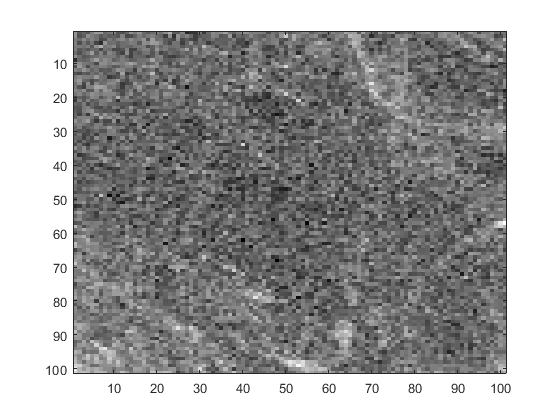}
\endminipage\hfill
\caption{Visual separation of the seven target blocks for $\alpha=0.1$: We exhibit the mean power in dB over the 186 bands. Columns from left to right: the original HSI containing the seven target blocks, low-rank background HSI $\mathbf{L}$, sparse target HSI $(\mathbf{A}_t \mathbf{C})^T$, and noise HSI.}
\label{fig:visual1}
\end{figure*}

\section{Experiments and Analysis}
\label{sec:experimentss}
In what follows, we perform both synthetic as well as real experiments to gauge the target detection performances of the two preceding strategies in Section \ref{sec:what_after}.
\\
The evaluations are done on two small zones acquired from a scene formed by a concatenation of two sectors labeled as ``f970619t01p02\_r02\_sc03.a.rf'' and
``f970619t01p02\_r02\_sc04.a.rfl'' in the online Cuprite HSI data \cite{CupriteHSIOnline}. The Cuprite HSI is a mining district area which is well understood mineralogically \cite{Swayze10245, JGRE:JGRE1642}. It contains well exposed zones of advanced argillic alteration, consisting principally of kiolinite, alunite, and hydrothermal silica. It was acquired by the Airborne Visible/Infrared Imaging Spectrometer (AVIRIS) in June 23, 1995 at local noon and under high visibility conditions by a NASAER-2 aircraft flying at an altitude of 20 km. It consists of 224 spectral (color) bands in contiguous (of about 0.01 $\mu$m) wavelengths ranging exactly from 0.4046 to 2.4573 $\mu$m. Prior to some analysis of the Cuprite HSI, the spectral bands 1-4, 104-113, 148-167, {\color{red}and 221-224} are removed due to the water absorption in those bands. As a result, a total of 186 bands are used.

\begin{figure*}[!tbp]
\minipage{0.2\textwidth}
  \includegraphics[width=\linewidth]{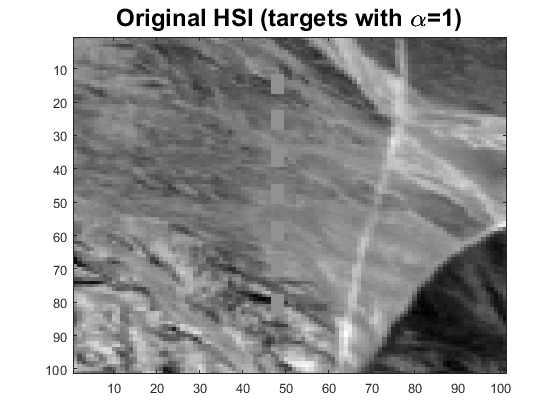}
\endminipage\hfill
\minipage{0.2\textwidth}
  \includegraphics[width=\linewidth]{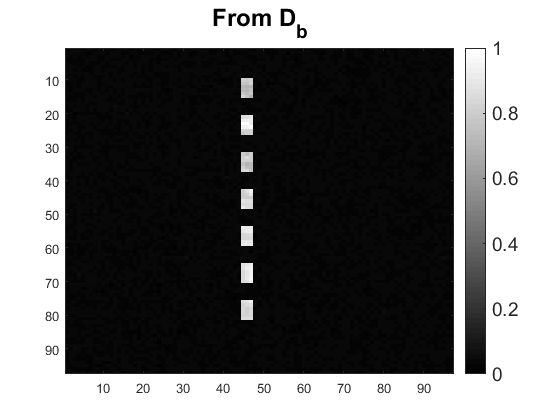}
\endminipage\hfill
\minipage{0.2\textwidth}
  \includegraphics[width=\linewidth]{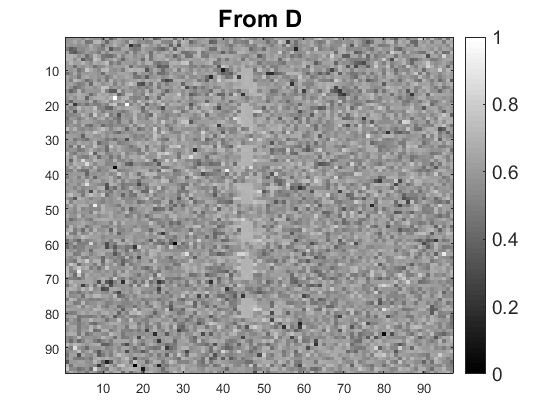}
\endminipage\hfill
\minipage{0.2\textwidth}
  \includegraphics[width=\linewidth]{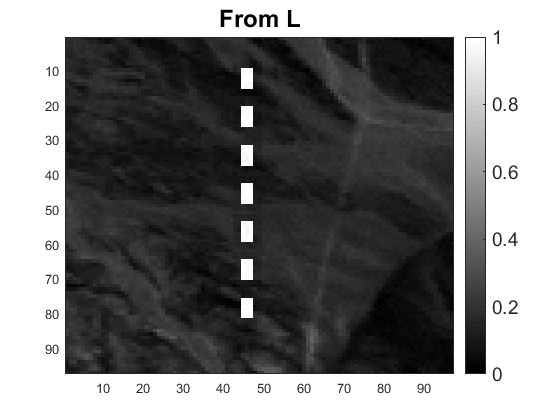}
\endminipage\hfill
\minipage{0.2\textwidth}
 \includegraphics[width=\linewidth]{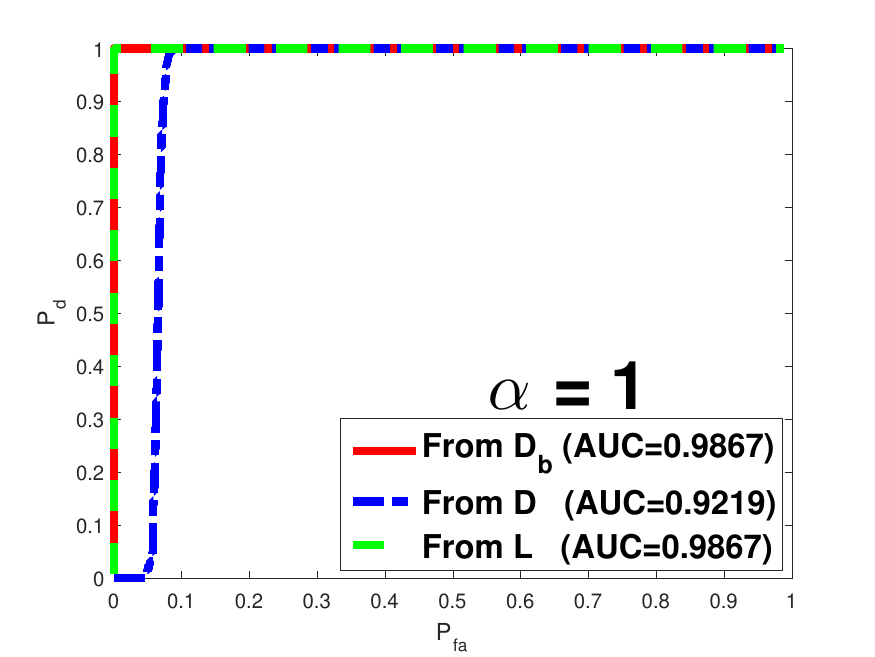}
\endminipage\hfill

\minipage{0.2\textwidth}
  \includegraphics[width=\linewidth]{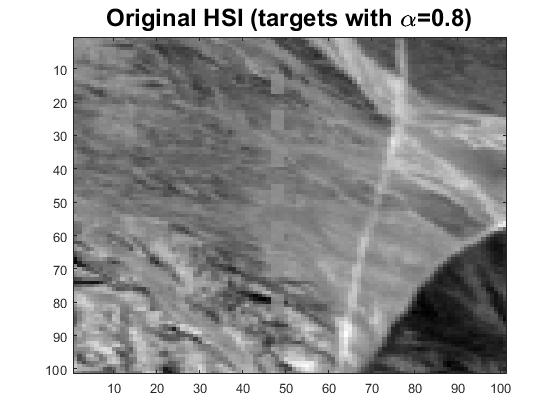}
\endminipage\hfill
\minipage{0.2\textwidth}
  \includegraphics[width=\linewidth]{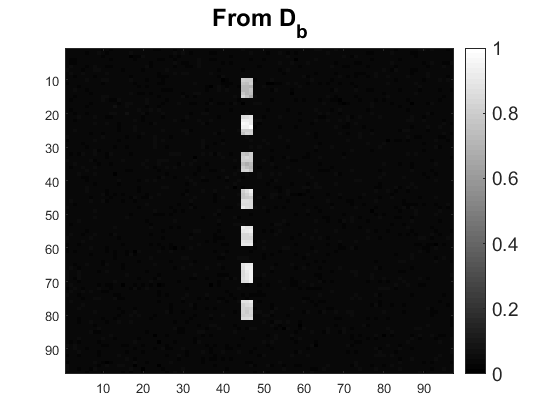}
\endminipage\hfill
\minipage{0.2\textwidth}
  \includegraphics[width=\linewidth]{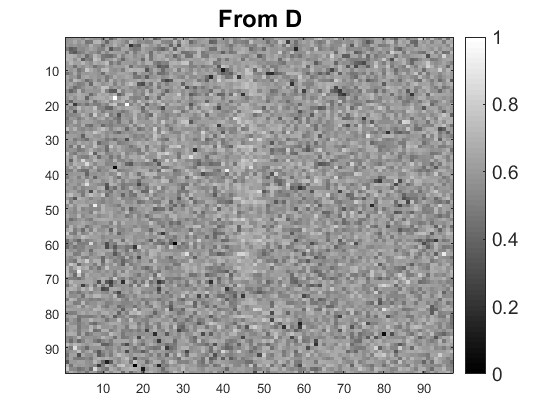}
\endminipage\hfill
\minipage{0.2\textwidth}
  \includegraphics[width=\linewidth]{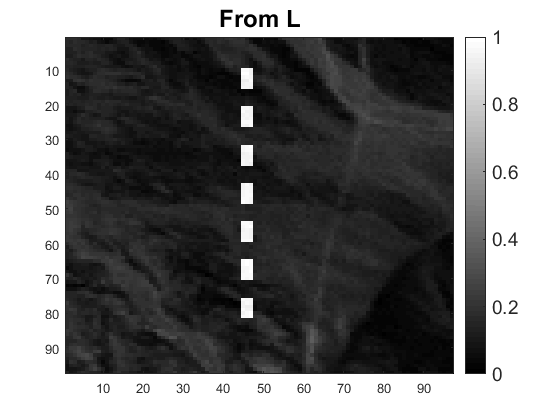}
\endminipage\hfill
\minipage{0.2\textwidth}
 \includegraphics[width=\linewidth]{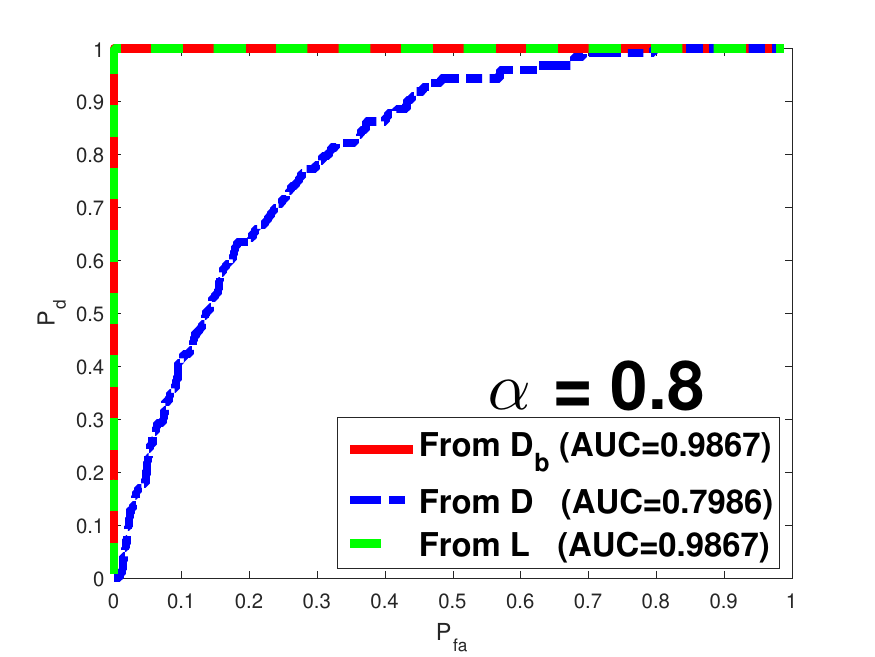}
\endminipage\hfill

\minipage{0.2\textwidth}
  \includegraphics[width=\linewidth]{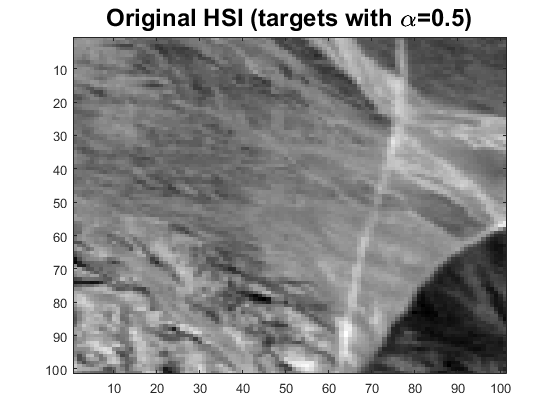}
\endminipage\hfill
\minipage{0.2\textwidth}
  \includegraphics[width=\linewidth]{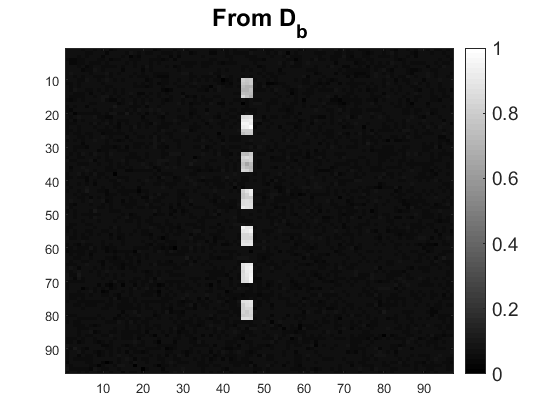}
\endminipage\hfill
\minipage{0.2\textwidth}
  \includegraphics[width=\linewidth]{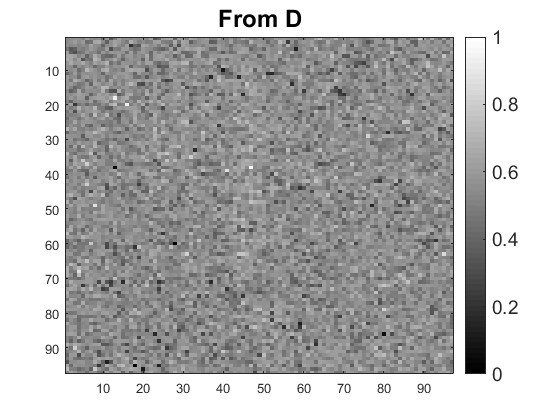}
\endminipage\hfill
\minipage{0.2\textwidth}
  \includegraphics[width=\linewidth]{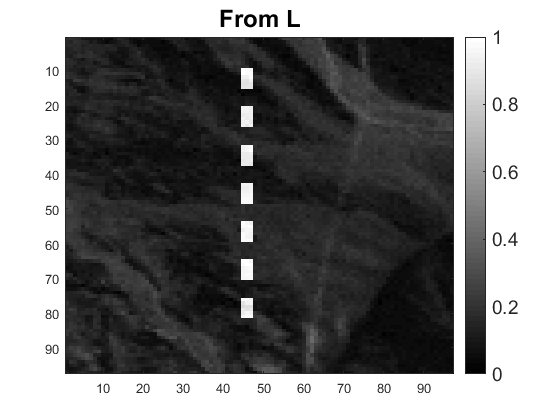}
\endminipage\hfill
\minipage{0.2\textwidth}
 \includegraphics[width=\linewidth]{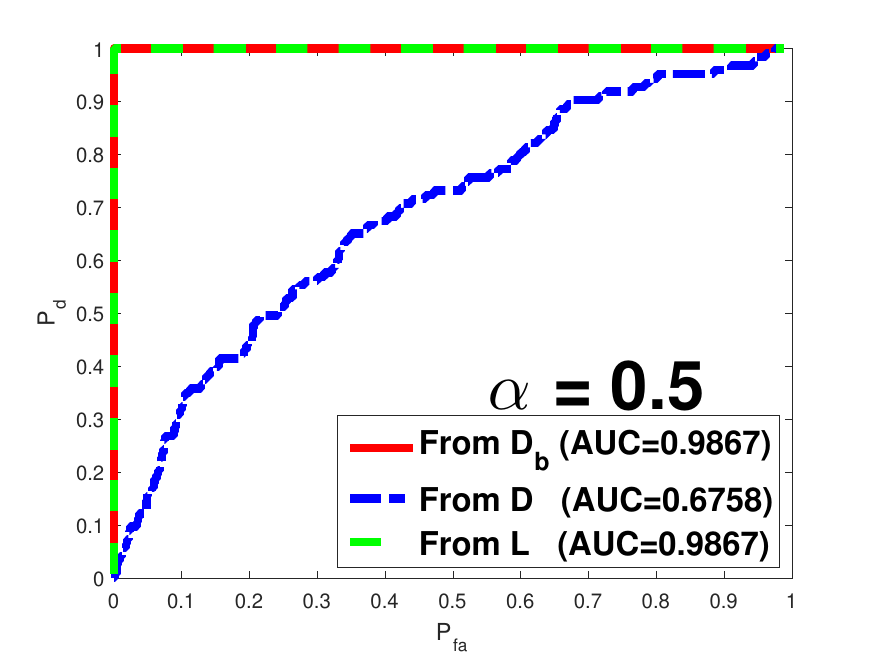}
\endminipage\hfill

\minipage{0.2\textwidth}
  \includegraphics[width=\linewidth]{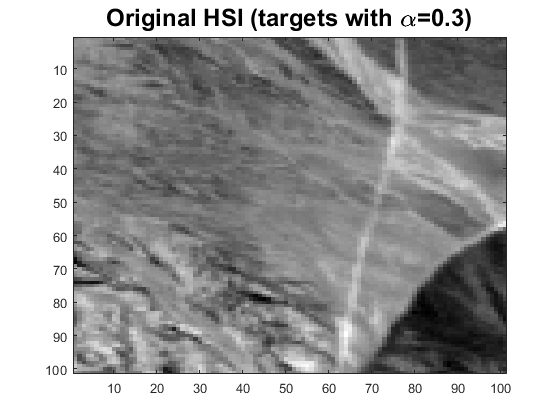}
\endminipage\hfill
\minipage{0.2\textwidth}
  \includegraphics[width=\linewidth]{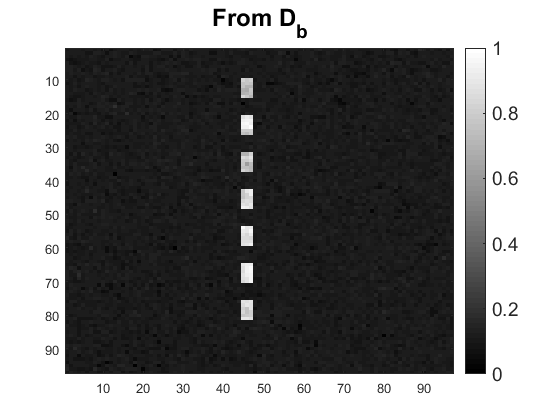}
\endminipage\hfill
\minipage{0.2\textwidth}
  \includegraphics[width=\linewidth]{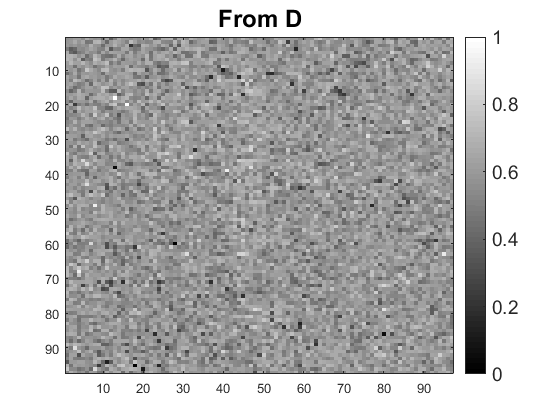}
\endminipage\hfill
\minipage{0.2\textwidth}
  \includegraphics[width=\linewidth]{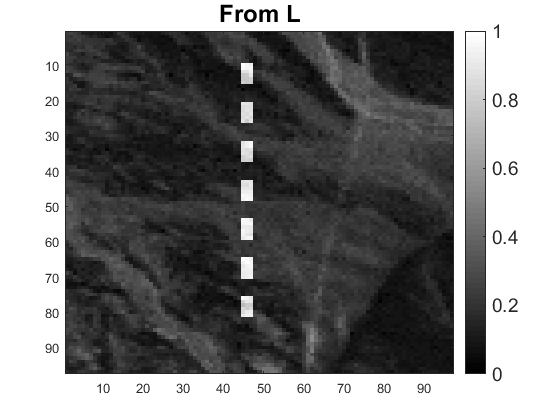}
\endminipage\hfill
\minipage{0.2\textwidth}
 \includegraphics[width=\linewidth]{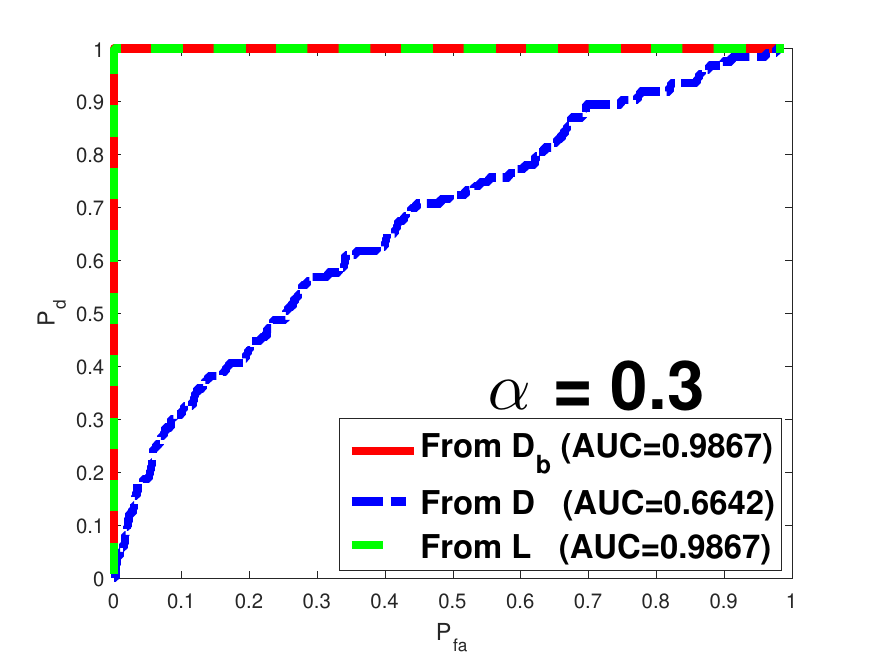}
\endminipage\hfill

\minipage{0.2\textwidth}
  \includegraphics[width=\linewidth]{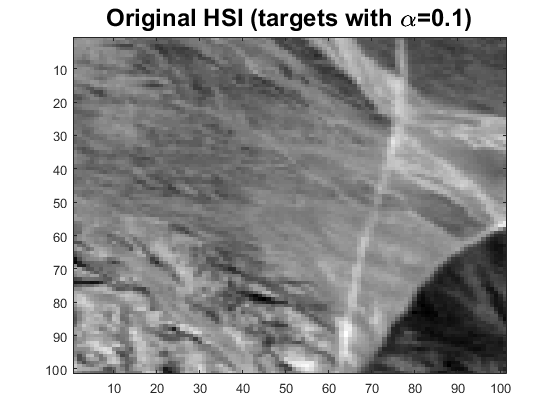}
\endminipage\hfill
\minipage{0.2\textwidth}
  \includegraphics[width=\linewidth]{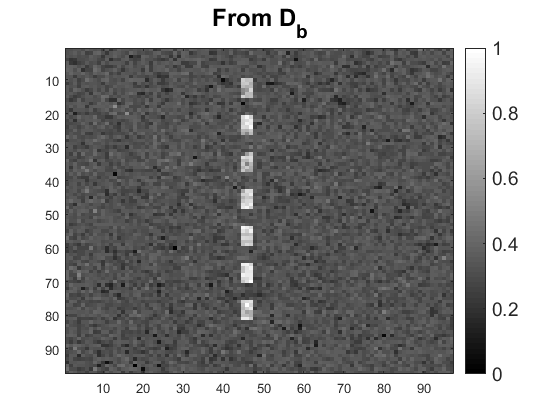}
\endminipage\hfill
\minipage{0.2\textwidth}
  \includegraphics[width=\linewidth]{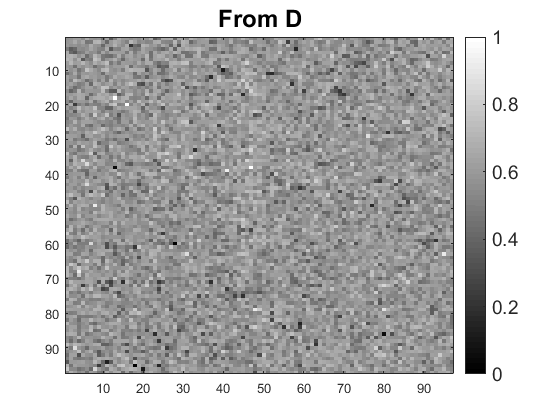}
\endminipage\hfill
\minipage{0.2\textwidth}
  \includegraphics[width=\linewidth]{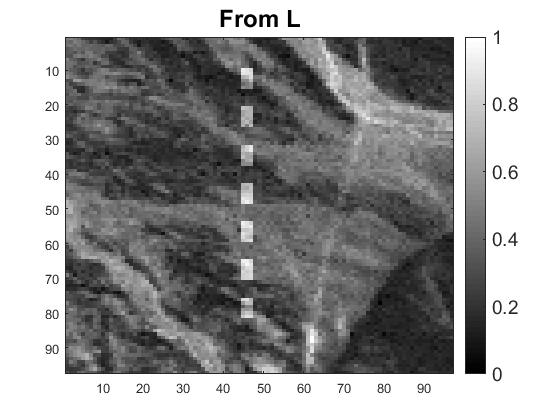}
\endminipage\hfill
\minipage{0.2\textwidth}
 \includegraphics[width=\linewidth]{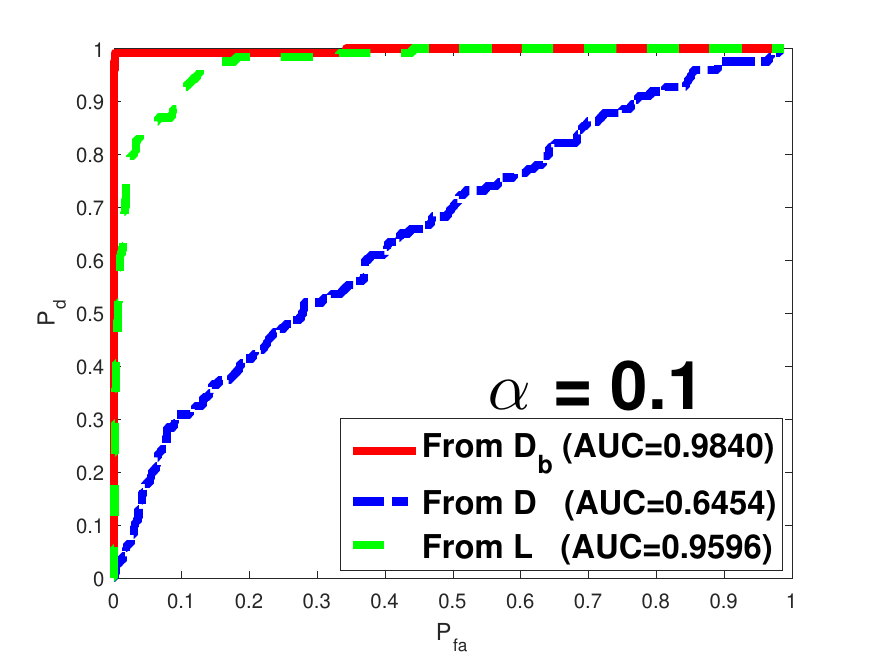}
\endminipage\hfill

\minipage{0.2\textwidth}
  \includegraphics[width=\linewidth]{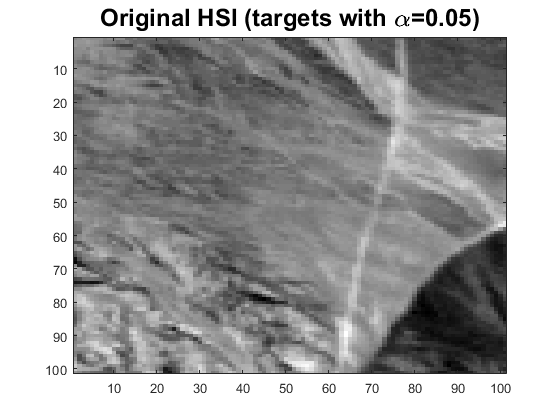}
\endminipage\hfill
\minipage{0.2\textwidth}
  \includegraphics[width=\linewidth]{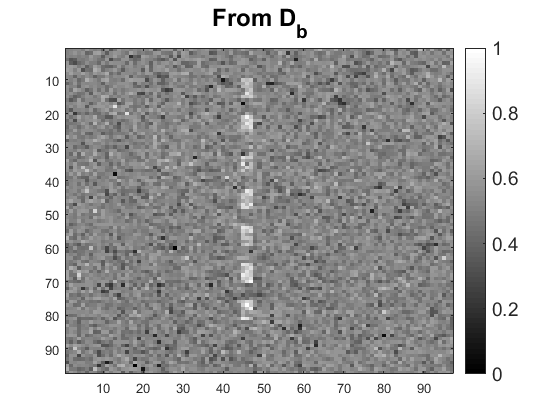}
\endminipage\hfill
\minipage{0.2\textwidth}
  \includegraphics[width=\linewidth]{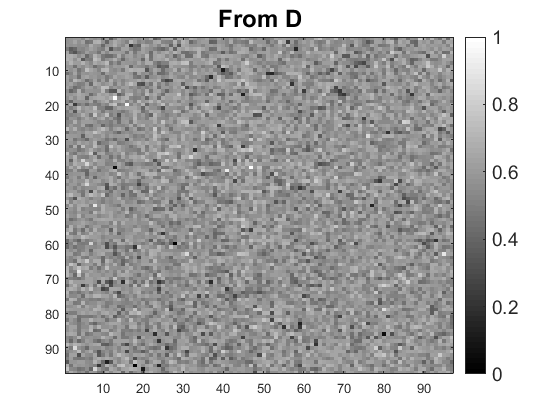}
\endminipage\hfill
\minipage{0.2\textwidth}
  \includegraphics[width=\linewidth]{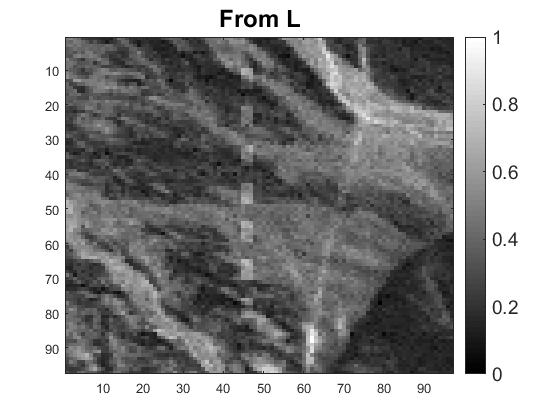}
\endminipage\hfill
\minipage{0.2\textwidth}
 \includegraphics[width=\linewidth]{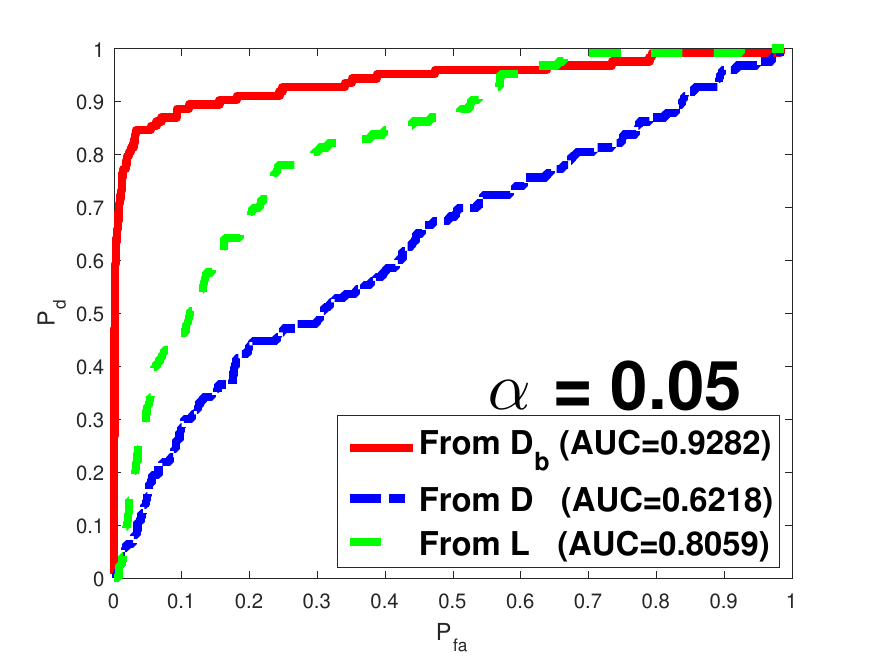}
\endminipage\hfill

\minipage{0.2\textwidth}
  \includegraphics[width=\linewidth]{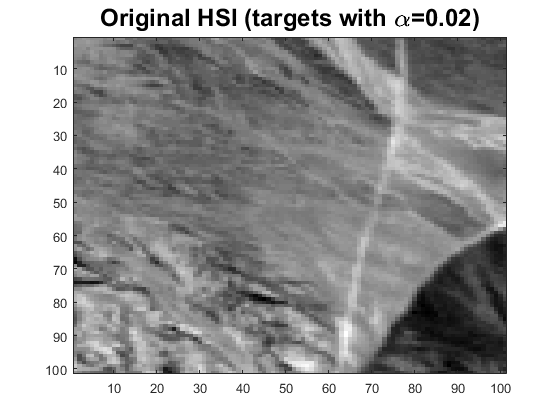}
\endminipage\hfill
\minipage{0.2\textwidth}
  \includegraphics[width=\linewidth]{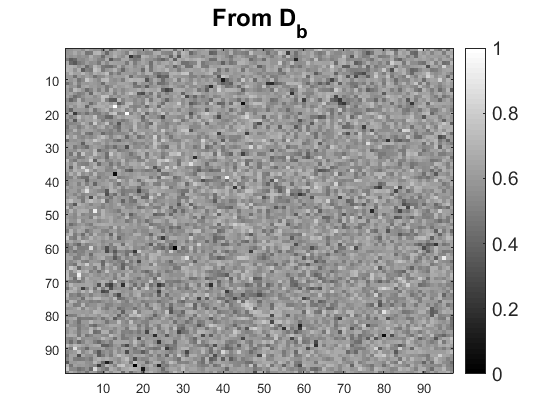}
\endminipage\hfill
\minipage{0.2\textwidth}
  \includegraphics[width=\linewidth]{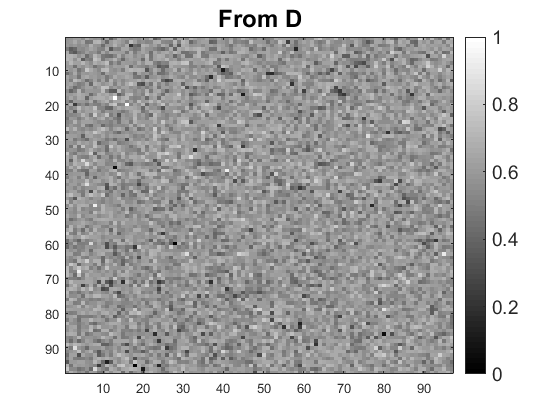}
\endminipage\hfill
\minipage{0.2\textwidth}
  \includegraphics[width=\linewidth]{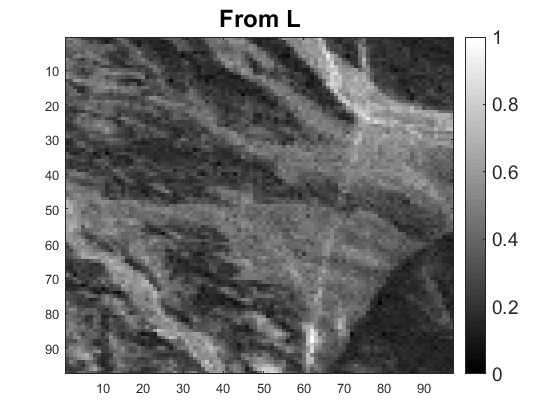}
\endminipage\hfill
\minipage{0.2\textwidth}
 \includegraphics[width=\linewidth]{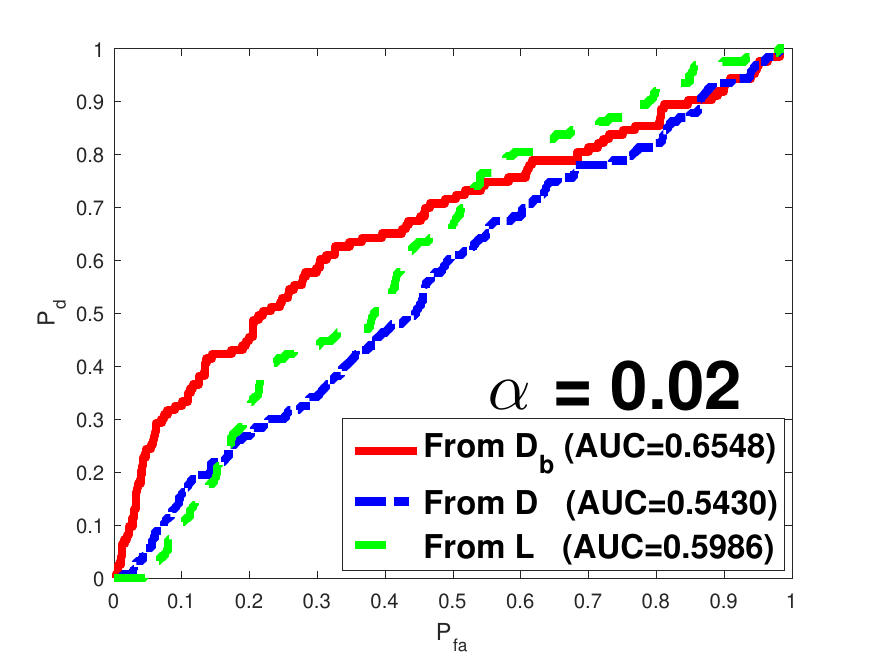}
\endminipage\hfill

\minipage{0.2\textwidth}
  \includegraphics[width=\linewidth]{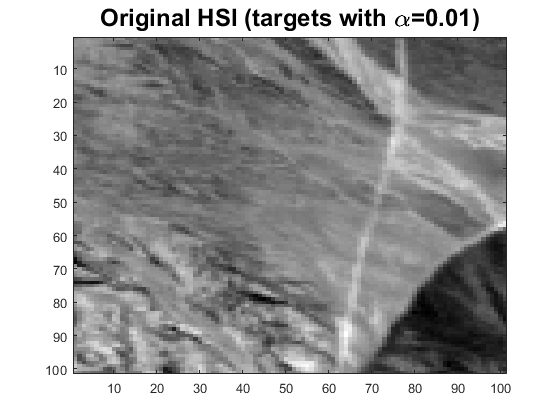}
\endminipage\hfill
\minipage{0.2\textwidth}
  \includegraphics[width=\linewidth]{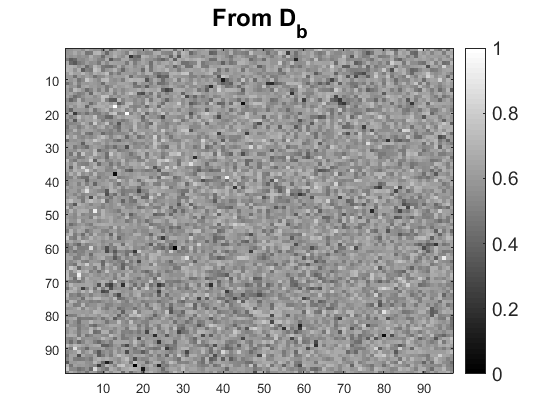}
\endminipage\hfill
\minipage{0.2\textwidth}
  \includegraphics[width=\linewidth]{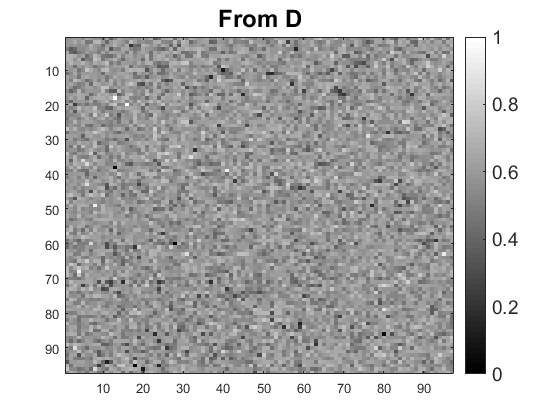}
\endminipage\hfill
\minipage{0.2\textwidth}
  \includegraphics[width=\linewidth]{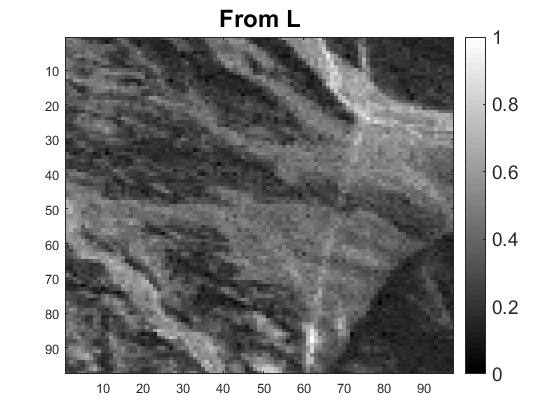}
\endminipage\hfill
\minipage{0.2\textwidth}
 \includegraphics[width=\linewidth]{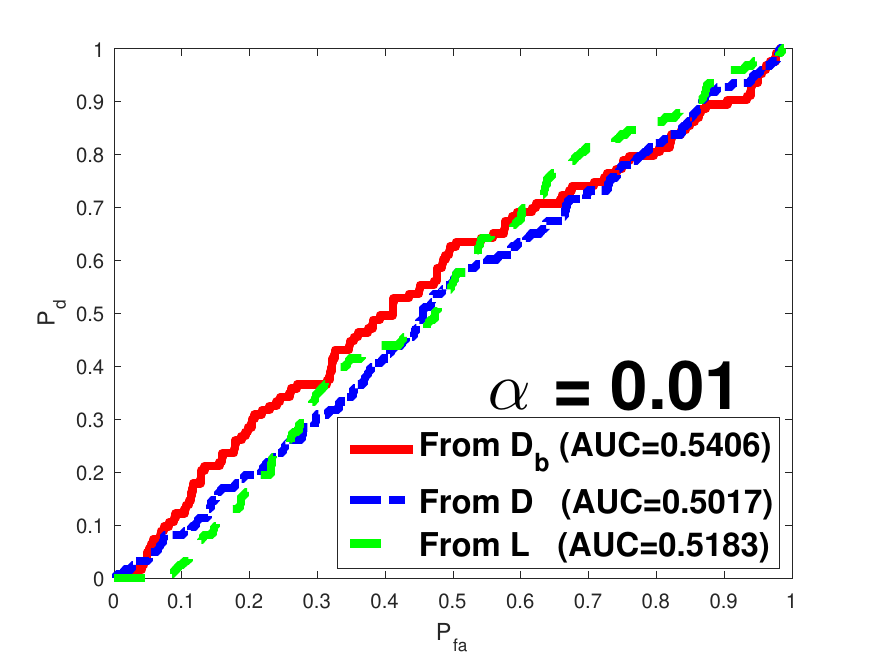}
\endminipage
\caption{Visual detection results and ROC curves [with their area under curve (AUC) values] for different $\alpha$ values of the SRBBH detector when $\mathbf{A}_b$ is constructed from $\mathbf{D}_b$, $\mathbf{D}$  and $\mathbf{L}$. Note that to have a fair comparison between the SRBBH outputs (to have the same color scales), we have normalized each of them to the values between 0 and 1.}
\label{fig:detec1}
\end{figure*}
\begin{figure*}[!tbp]
\minipage{0.25\textwidth}
  \includegraphics[width=\linewidth]{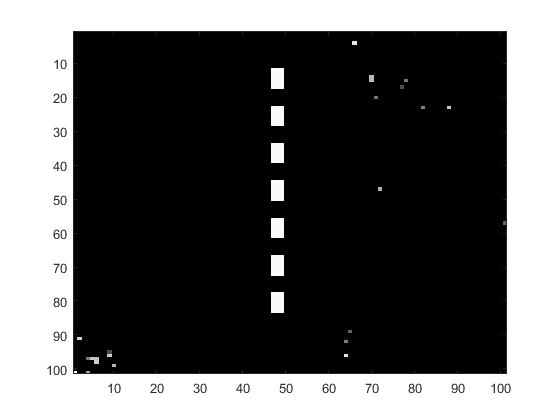}
{\begin{center}%
\vspace{-5mm}
{$\bf \alpha=1$}
\end{center}}
\endminipage\hfill
\minipage{0.25\textwidth}
  \includegraphics[width=\linewidth]{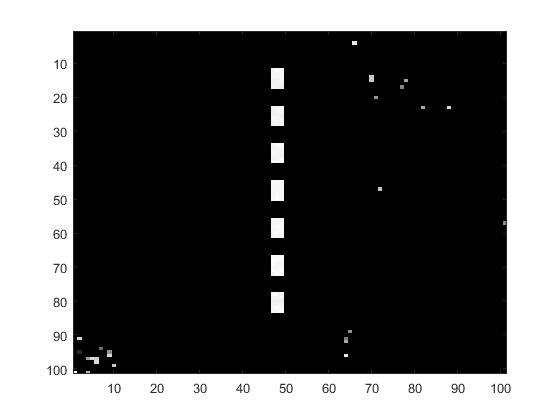}
{\begin{center}%
\vspace{-5mm}
{$\bf \alpha=0.8$}
\end{center}}
\endminipage\hfill
\minipage{0.25\textwidth}
  \includegraphics[width=\linewidth]{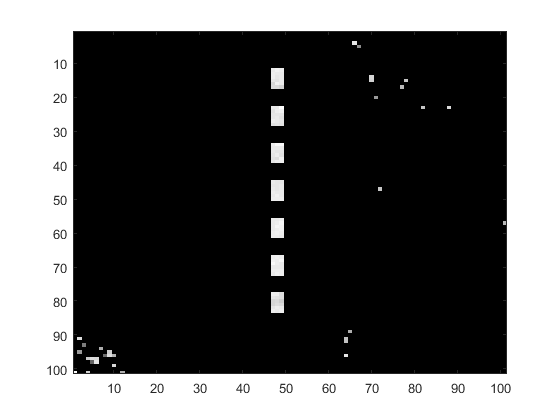}
{\begin{center}%
\vspace{-5mm}
{$\bf \alpha=0.5$}
\end{center}}
\endminipage\hfill
\minipage{0.25\textwidth}
  \includegraphics[width=\linewidth]{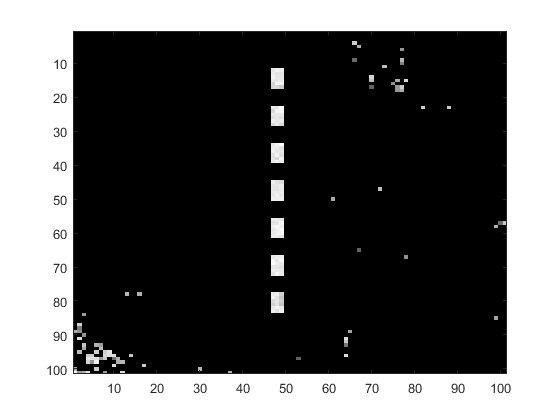}
{\begin{center}%
\vspace{-5mm}
{$\bf \alpha=0.3$}
\end{center}}
\endminipage\hfill
\minipage{0.25\textwidth}
  \includegraphics[width=\linewidth]{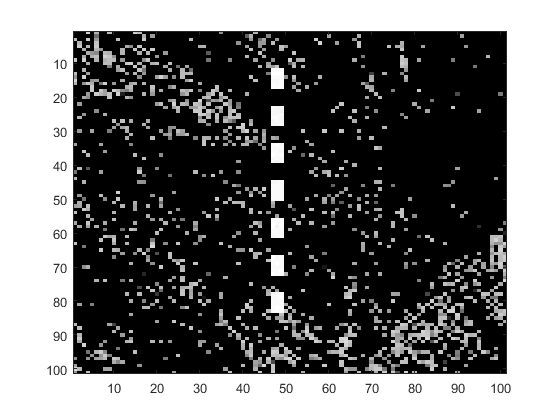}
{\begin{center}%
\vspace{-5mm}
{$\bf \alpha=0.1$}
\end{center}}
\endminipage\hfill
\minipage{0.25\textwidth}
  \includegraphics[width=\linewidth]{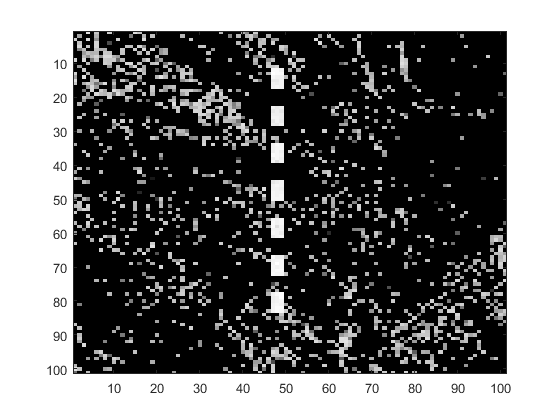}
{\begin{center}%
\vspace{-5mm}
{$\bf \alpha=0.05$}
\end{center}}
\endminipage\hfill
\minipage{0.25\textwidth}
  \includegraphics[width=\linewidth]{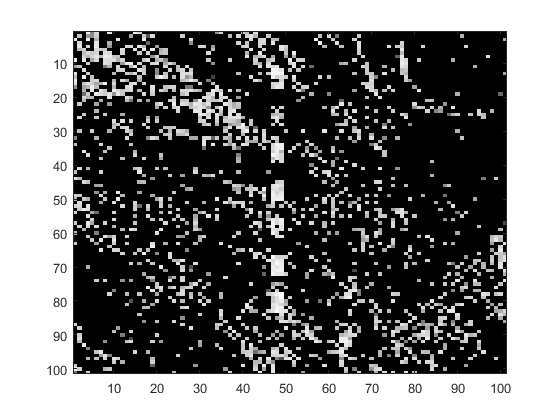}
{\begin{center}%
\vspace{-5mm}
{$\bf \alpha=0.02$}
\end{center}}
\endminipage\hfill
\minipage{0.25\textwidth}
  \includegraphics[width=\linewidth]{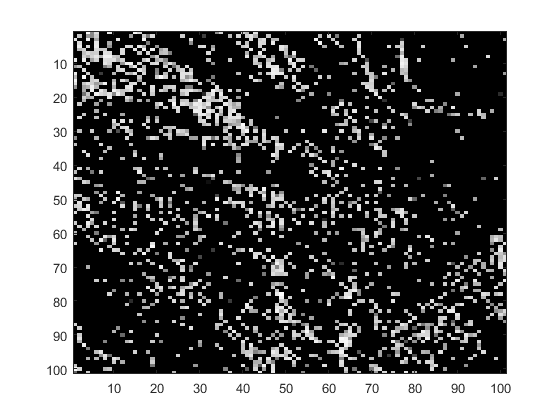}
{\begin{center}%
\vspace{-5mm}
{$\bf \alpha=0.01$}
\end{center}}
\endminipage
\caption{Visual detections (mean power in dB over the 186 bands) of $\left(\mathbf{A}_t\mathbf{C}\right)^T$ for the seven target blocks for different $\alpha$ values. (From top-left to bottom-right) Values of $\alpha$: 1, 0.8, 0.5, 0.3, 0.1, 0.05, 0.02, and 0.01.}
\label{fig:visual3}
\end{figure*}

\begin{figure*}[!tbp]
\minipage{0.25\textwidth}
  \includegraphics[width=\linewidth]{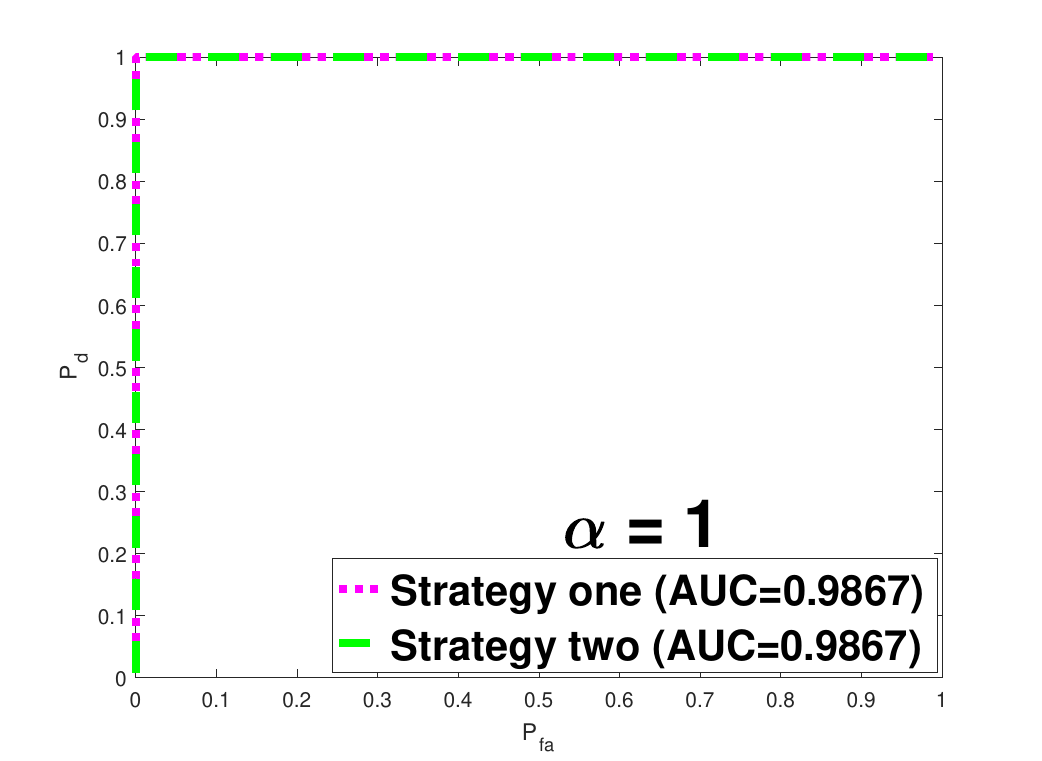}
\endminipage\hfill
\minipage{0.25\textwidth}
  \includegraphics[width=\linewidth]{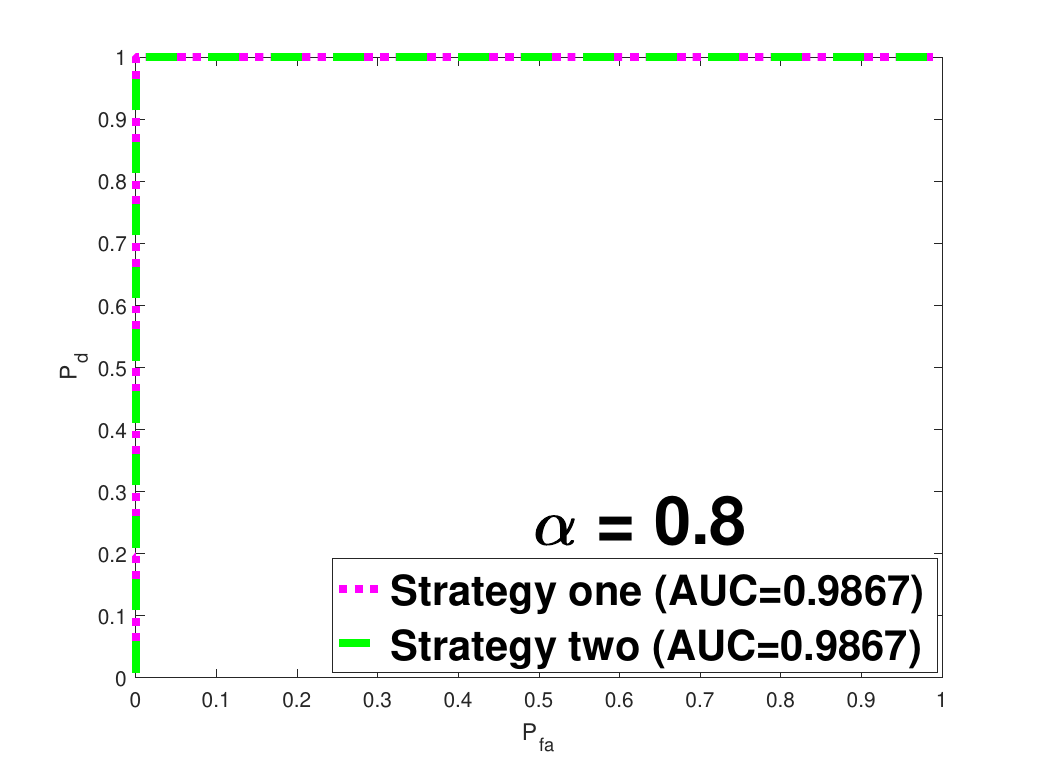}
\endminipage\hfill
\minipage{0.25\textwidth}
  \includegraphics[width=\linewidth]{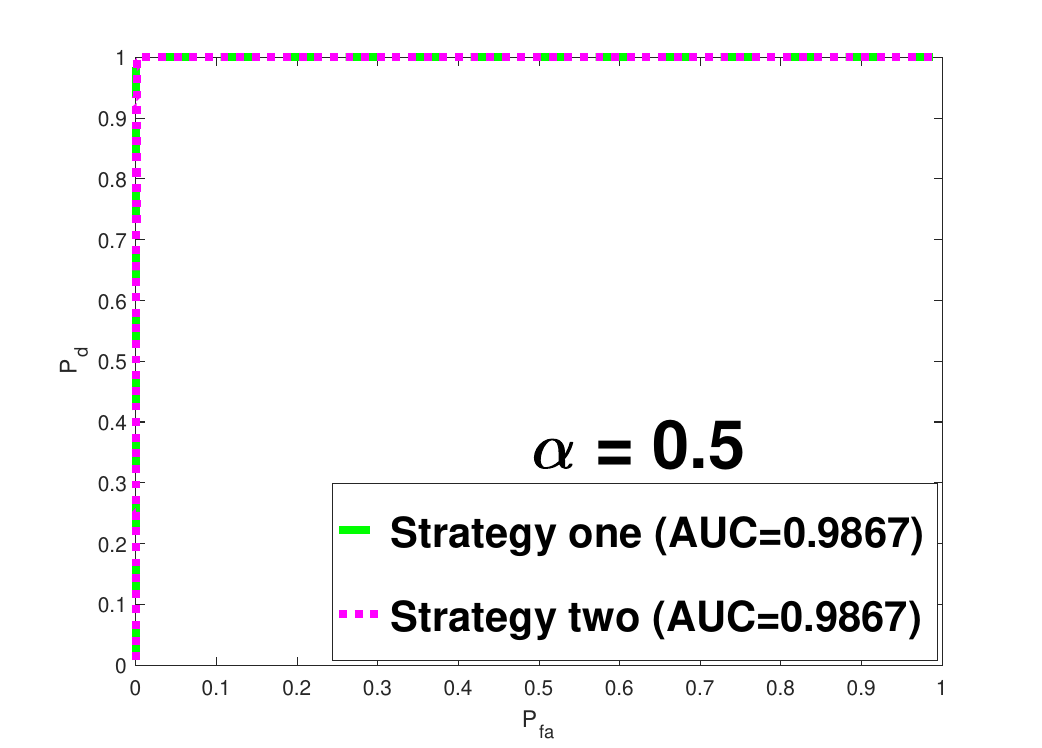}
\endminipage\hfill
\minipage{0.25\textwidth}
  \includegraphics[width=\linewidth]{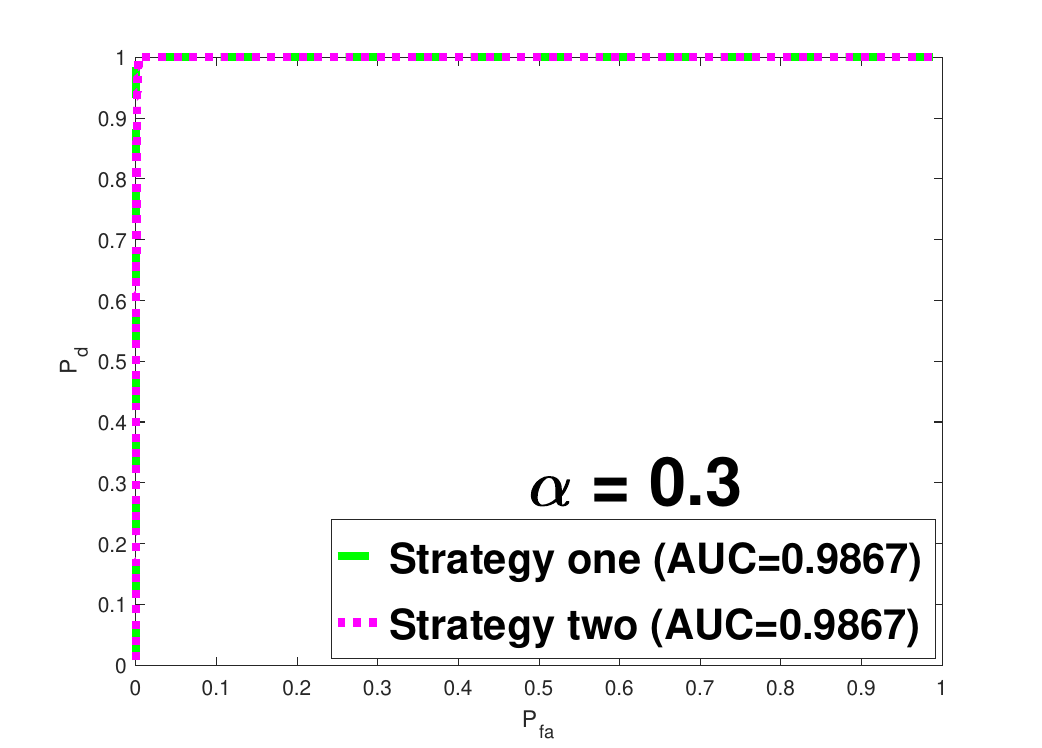}
\endminipage\hfill
\minipage{0.25\textwidth}
  \includegraphics[width=\linewidth]{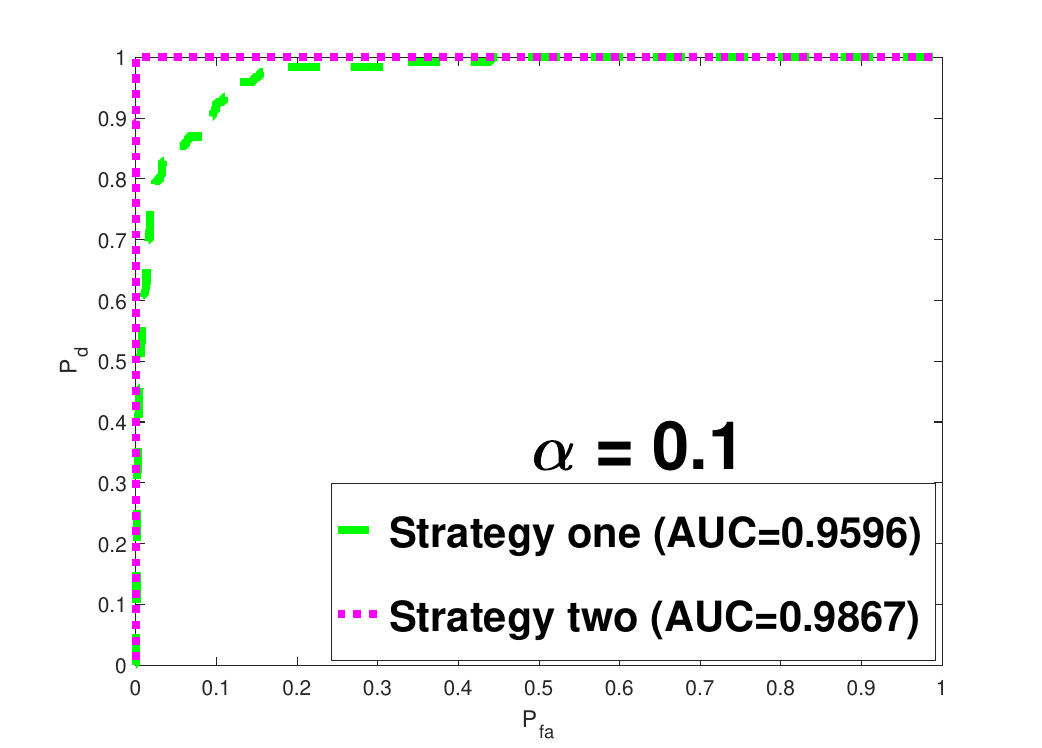}
\endminipage\hfill
\minipage{0.25\textwidth}
  \includegraphics[width=\linewidth]{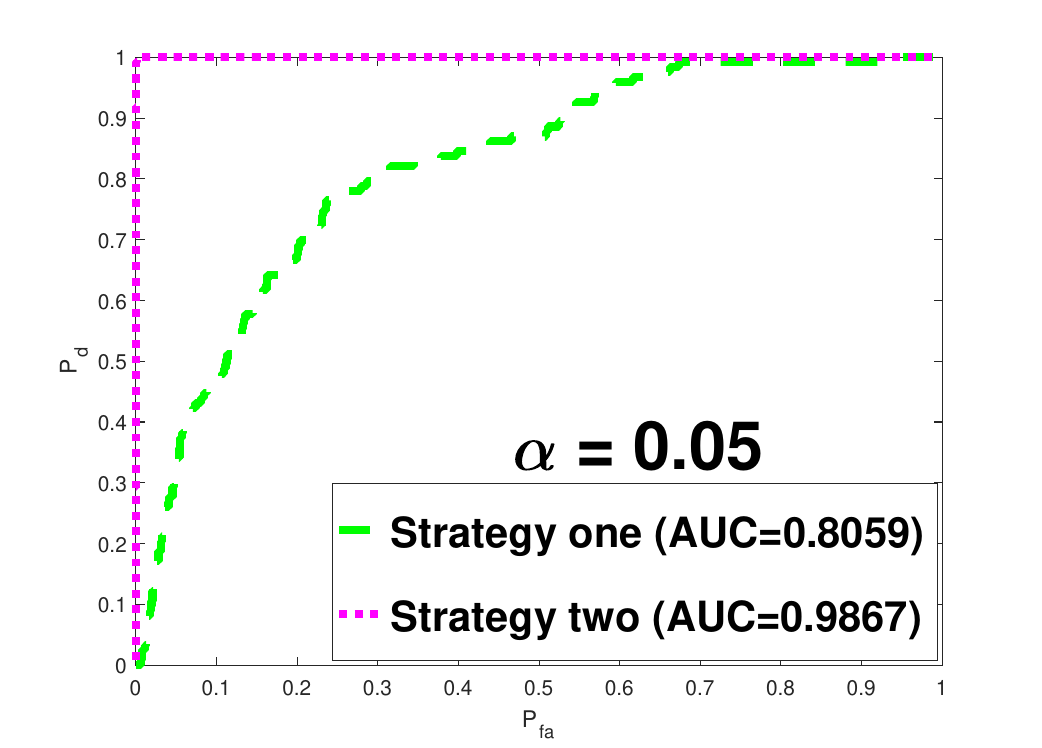}
\endminipage\hfill
\minipage{0.25\textwidth}
  \includegraphics[width=\linewidth]{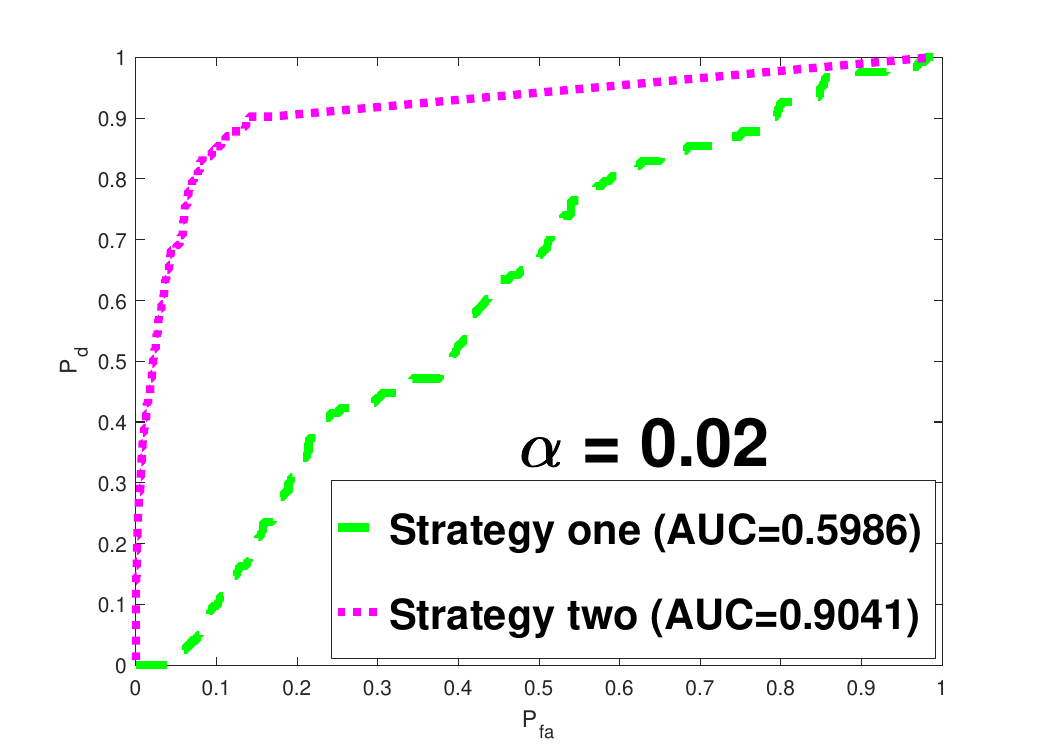}
\endminipage\hfill
\minipage{0.25\textwidth}
  \includegraphics[width=\linewidth]{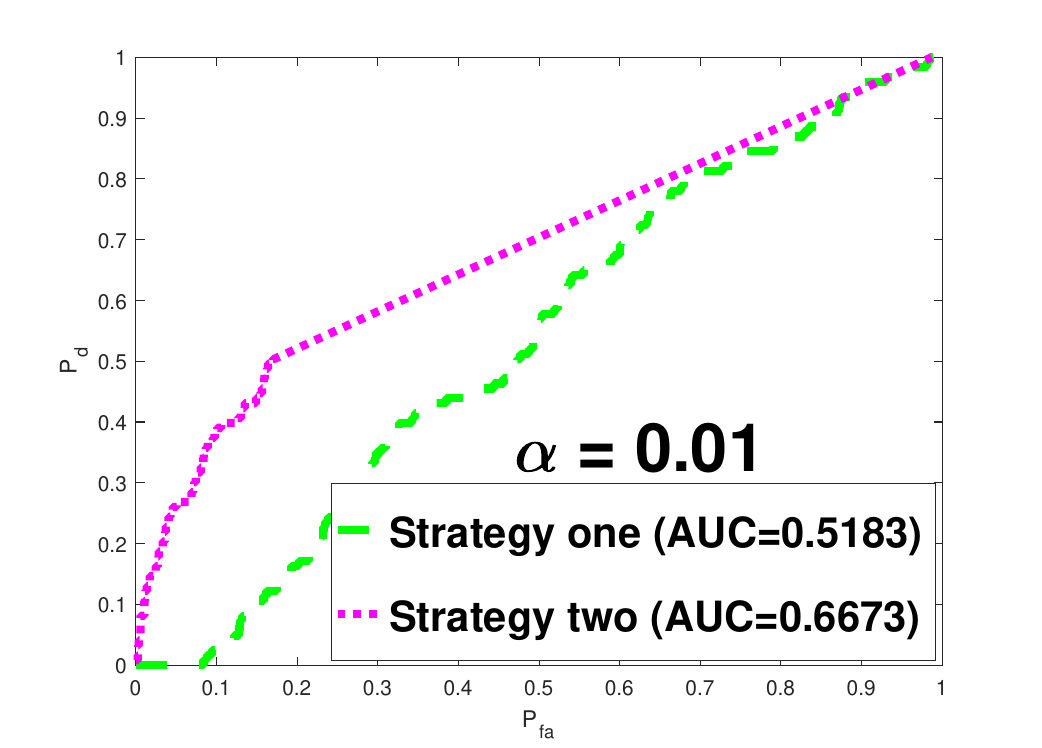}
\endminipage
\caption{Detection comparisons between Strategy one and Strategy two.}
\label{fig:detec7}
\end{figure*}
\subsection{General Discussion About the Parameters $\tau$ and $\lambda$}
In this section, we discuss the main difficulties that face our problem \eqref{eq:convex_model} in accurately choosing the values of $\tau$ and $\lambda$ for both Strategy one and Strategy two.
Currently, $\tau$ and $\lambda$ are set manually (to achieve the best target detection performance) for both the strategies. However, this manual selection depends on the HSI used, on the spatial and spectral dimensions of the given HSI, on the targets present, and even on how accurate the target dictionary $\mathbf{A}_t$ is. All these challenges strongly encourage us to alleviate the manual selection problem of $\tau$ and $\lambda$ by an automatic method in the future.
\begin{center}
{\em How can we play now with $\tau$ and $\lambda$ ?}
\end{center}
We found that a useful way to set $\tau$ and $\lambda$ would be to decide on the ratio of $\tau$ and $\lambda$, respectively, for both the strategies and then set the relative values of the weights between the first two terms and the third term in \eqref{eq:convex_model}.

For Strategy two, we found that the ratio of $\tau$ to $\lambda$ must be equal to $\frac{5}{2}$ in both synthetic and real experiments. For Strategy one, we found that the ratios of $\tau$ to $\lambda$ should be high to make sure that all of the targets are removed to the target image. We set this ratio to approximately $6$ for the synthetic experiments and $10$ for the real experiments. The ratio for the latter case must be higher because for the real experiments, we do not really have comprehensive enough target dictionaries to represent the target well, and thus, we need extra incentive for the target fractions to go to the target image. 

The different requirements imposed by the two strategies that can lead to our particular choice setting of the $\tau$ to $\lambda$ ratio also dictate how we should set the relative values of the weights between the first two terms and the third term in \eqref{eq:convex_model} as follows.
\begin{enumerate}
\item A lower penalty associated with the third term (that is, by raising the absolute levels of $\tau$ and $\lambda$) would tolerate more deviation and thus encourage more noise or image clutters (by image clutters, we mean the small heterogeneous objects and specular highlights) to be absorbed by this term. This is particularly important for Strategy two when there are a lot of image clutters that do not exactly conform to a low-rank background model: since these clutters do not satisfy the low-rank property, they have a propensity to show up in the second term if we do not sufficiently lower the penalty for the third term and, thus, contribute to a lot of false alarms for Strategy two.

\item On the other hand, such a low-penalty setting for the third term may not be a good idea for Strategy one, as the third term absorbs too much of the image clutters that actually form the background, causing the background dictionary $\mathbf{A}_b$ so constructed to lose representative power.
\end{enumerate}
In summary, for Strategy two, we set $\tau$ and $\lambda$ at $0.05$ and $0.02$ in the synthetic experiments, whereas at $0.5$ and $0.2$ in the real experiments.  For Strategy one, we set $\tau$ and $\lambda$ at $0.8$ and $0.133$ in the synthetic experiments, whereas at $3$ and $0.3$ in the real experiments.

\subsection{Synthetic Experiments}
\label{sec:synthetic_Experiments}
The experiments are done on a $101\times101$ zone (pixels in rows 389$-$489 and columns 379$-$479) from the acquired cuprite scene.
We incorporate, in this zone, seven target blocks (each of size $6\times 3$) with $\alpha\in[0.01, \, 1]$ (all have the same $\alpha$ value), placed in long convoy formation all formed by the same synthetic (perfect) target $\mathbf{t}$ consisting of a sulfate mineral type known as ``jarosite''. We make sure by referring to Fig. 5(a) in \cite{Swayze10245} that the small zone we consider does not already contain any jarosite patches.
The target $\mathbf{t}$ that we created actually consists of the mean of the first six jarosite mineral samples taken from the online United States Geological Survey (USGS - Reston) spectral library \cite{Clark93} (see Fig. \ref{fig:Jarosite_target_samples}). The target $\mathbf{t}$ replaces a fraction $\alpha\in[0.01, \, 1]$ from the background; specifically, the following values of $\alpha$ are considered: 0.01, 0.02, 0.05, 0.1, 0.3, 0.5, 0.8, and 1. As for the target dictionary $\mathbf{A}_t$, it is constructed from the six acquired jarosite samples\footnote{Note that both the HSI and the jarosite target samples are normalized to the values between 0 and 1.}.  
\\

\subsubsection{Using Strategy One for Detection}
\label{sec:Strategy_one_synthetic}

We first provide, in Fig. \ref{fig:visual1}, a visual evaluation of the separation of the above-mentioned seven target examples for low $\alpha=0.1$. We can observe that our problem \eqref{eq:convex_model} successfully discriminates these perceptually invisible targets from the background in $\mathbf{D}$ and separate them. The seven darker blocks that appear in $\mathbf{L}$ correspond to the dimmer fraction of the background that remains after the targets have been removed at the corresponding spatial locations.

Having qualitatively inspected the separation, we, now, aim to qualitatively and quantitatively evaluate the target detection performances of the SRBBH detector \cite{Zhang15} when $\mathbf{A}_b$ is constructed using a small concentric window of size $5\times 5$. That is, $\mathbf{A}_b\in\mathbb{R}^{p\times24}$ (after excluding the center pixel) and the region tested consists of an image of size $97 \times 97$.
\\
In what follows, we shall use $\mathbf{D}_b$ to represent the HSI that does not contain the seven target blocks (that is, the pure background image) and $\mathbf{D}$ to represent the HSI after incorporating the seven target blocks (that is, it contains the targets) for $\alpha\in[0.01, \, 1]$.
\\
We, now, consider the following three scenarios to form the columns in $\mathbf{A}_b$.
\begin{enumerate}
\item For each test pixel in $\mathbf{D}$, we create the concentric window on $\mathbf{D}_b$. This represents the ideal case since $\mathbf{A}_b$ is free from the targets.
\item For each test pixel in $\mathbf{D}$, we create the concentric window on $\mathbf{D}$.
\item For each test pixel in $\mathbf{D}$, we create the concentric window on the low-rank background HSI $\mathbf{L}$.
\end{enumerate}
The target detection performances are evaluated qualitatively as well as quantitatively specifically by the receiver operating characteristics (ROC) curves, which describe the probability of detection ($P_d$) against the probability of false alarm ($P_{fa}$), as we vary the threshold $\eta$ between the minimal and maximal values of each detector output. A good detector presents high $P_d$ values at low $P_{fa}$, i.e., the curve is closer to the top-left corner. More particularly, $P_d$ can be determined as the ratio of the number of the target pixels determined as target (that is, the detector output at each pixel on the target region exceeds the threshold value) and the total number of true target pixels, whereas $P_{fa}$ can be calculated by the ratio of the number of false alarms (the detector output at each pixel on the background region that is outside the target region exceeds the threshold value) and the total number of pixels in the region tested.
\\
Fig. \ref{fig:detec1} shows both the qualitative and quantitative detection results. Clearly, increasing $\alpha$ should render the target detection less challenging, and thus, better detection results are being expected. However, this fact cannot always be the case for the SRBBH detector when $\mathbf{A}_b$ is constructed from $\mathbf{D}$ (blue curve): it is true that the increase in $\alpha$ helps to improve the detection, but, at the same time, leads to more target contamination in $\mathbf{A}_b$, which in turn suppresses the detection improvement that ought to be had. That is why the SRBBH detector (blue curve) does not reap full benefits from the increase in $\alpha$ and, thus, presents the poor detection results even for large $\alpha$ values. 
\\
By constructing $\mathbf{A}_b$ from $\mathbf{L}$, which only contains the background with the targets removed after applying problem \eqref{eq:convex_model}, the SRBBH detector (green dashed curve) can better detect the targets, especially for $\alpha\geq 0.1$, and has competitive detection results compared with the ideal case when $\mathbf{A}_b$ is constructed from $\mathbf{D}_b$. The detection performances start to deteriorate progressively for very small $\alpha$ values and degenerate to the SRBBH level (blue curve) for $\alpha\leq0.02$. 
\\
To sum up, the obtained target detection results corroborate our claim that we can handle targets with low fill-fraction and in convoy formation.
$\\$

\subsubsection{Using Strategy Two for Detection}
\label{synthetic_Strategytwo}
Fig. \ref{fig:visual3} shows the detection results of $\left(\mathbf{A}_t\, \mathbf{C}\right)^T$ for different $\alpha$ values. The plots correspond to the mean power in dB over the 186 spectral bands.
As can be seen, Strategy two detects all the targets with little false alarms until $\alpha \leq0.1$ when a lot of false alarms appear.
\\

\subsubsection{Further Discussion on the Obtained Detection Results for Both Strategy One and Strategy Two}
It is clear from the results of 2) and the preceding experiment that the value of the target fill-fraction impacts the performance substantially.
\\
For Strategy two, this is due to the relaxation of the $l_{2,0}$ norm to $l_{2,1}$ norm for the $\left(\mathbf{A}_t\, \mathbf{C}\right)^T$ term. Instead of counting the number of nonzero terms in $\left(\mathbf{A}_t\mathbf{C}\right)^T$, the magnitudes of these nonzero terms play a role too in the relaxed version. When the magnitudes of the target signals are small (for small $\alpha$), the penalty cost suffered is less. It follows that there is room for nontarget signals to appear or even take over in the $\left(\mathbf{A}_t\, \mathbf{C}\right)^T$ matrix, resulting in high false alarms and high miss rates.
\\
For Strategy one, this is not only due to the relaxation of the $l_{2,0}$ norm to $l_{2,1}$ norm for the $\left(\mathbf{A}_t\, \mathbf{C}\right)^T$ term, but also could be due to the approximation in solving the $l_0$ problem of $\boldsymbol{\theta}$ and $\boldsymbol{\gamma}$ in the SRBBH detector in \eqref{SRBBH} (here, the greedy method was used). For example, in Fig. \ref{fig:detec1}, for $\alpha\leq0.1$, the green curve has a lower AUC value than that of the red curve (the ideal one). This is mainly because of the $l_{2,1}$ relaxation in problem \eqref{eq:convex_model}. However, we can also notice how the detection of the red curve starts to decrease when $\alpha\leq0.1$ and degenerates to the blue curve for very small $\alpha$. This could be because of the $l_0$ approximation in solving $\boldsymbol{\theta}$ and $\boldsymbol{\gamma}$ in the SRBBH detector in \eqref{SRBBH}. 
\\

\subsubsection{Strategy One Versus Strategy Two: Detection Comparison}
The detection comparisons between Strategy one and Strategy two are done quantitatively via the ROC curves.
In order to be able to analyse the performances (that is, drawing the ROC curves) of Strategy two, we have proposed the following detector to be applied on each test pixel in $\left(\mathbf{A}_t\,\mathbf{C}\right)^T$:
\begin{equation}
D_{\left(\mathbf{A}_t\,\mathbf{C}\right)^T} (\mathbf{x}_s) = \frac{\mathbf{t}^T \,\mathbf{x}_s}{\mathbf{t}^T \, \mathbf{t}} \supinf \eta\, ,
\end{equation}
where $\mathbf{x}_s$ is being the test pixel in $\left(\mathbf{A}_t\mathbf{C}\right)^T$ and $\eta$ is the decision threshold to yield the desired probability of false alarm $P_{fa}$.  
We confirm that the visual detection results of the detector $D_{\left(\mathbf{A}_t\mathbf{C}\right)^T}$ are totally the same as to those in Fig. \ref{fig:visual3} for all values of $\alpha$. This has encouraged us to use $D_{\left(\mathbf{A}_t\,\mathbf{C}\right)^T}$ to plot the ROC curves for Strategy two.
\\
To plot the ROC curves for Strategy one, we have used the same small concentric window of size $5 \times 5$ (that is, $\mathbf{A}_b \in \mathbb{R}^{p \times 24}$).
\\
Fig. \ref{fig:detec7} shows the detection comparison results of both the strategies. Obviously, Strategy two achieves better detection results than to those of Strategy one, especially for small values of $\alpha$.

\begin{figure}[!tbp]
\centering
\minipage{0.4\textwidth}
  \includegraphics[width=\linewidth]{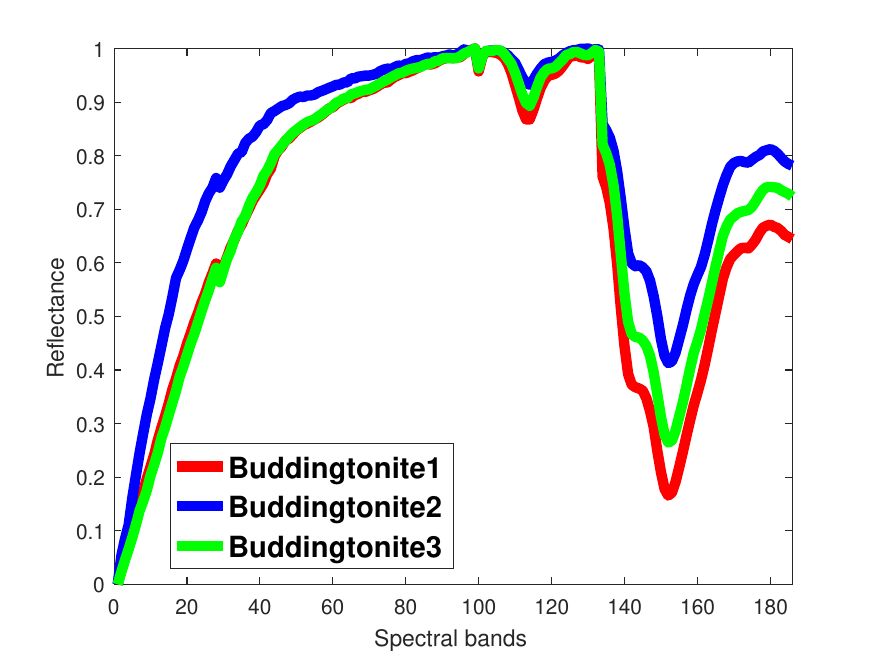}
\endminipage
 \caption{Plot of the buddingtonite target samples taken from the online ASTER spectral library.}
\label{fig:Buddingtonites_samples}
\end{figure}

\begin{figure*}[!tbp]
\minipage{0.25\textwidth}
  \includegraphics[width=\linewidth]{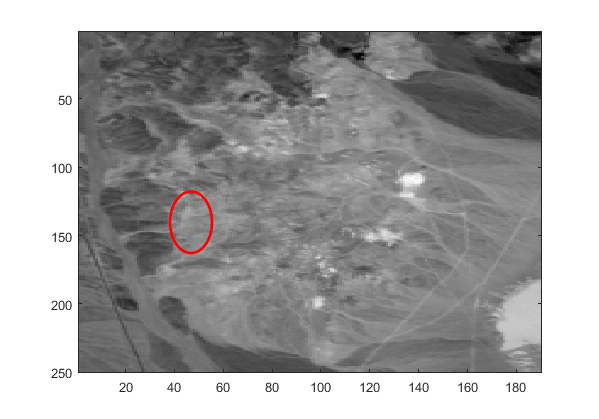}
\endminipage\hfill
\minipage{0.25\textwidth}
  \includegraphics[width=\linewidth]{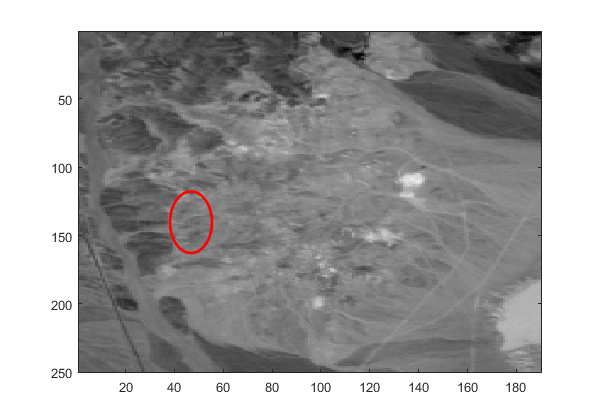}
\endminipage\hfill
\minipage{0.25\textwidth}
  \includegraphics[width=\linewidth]{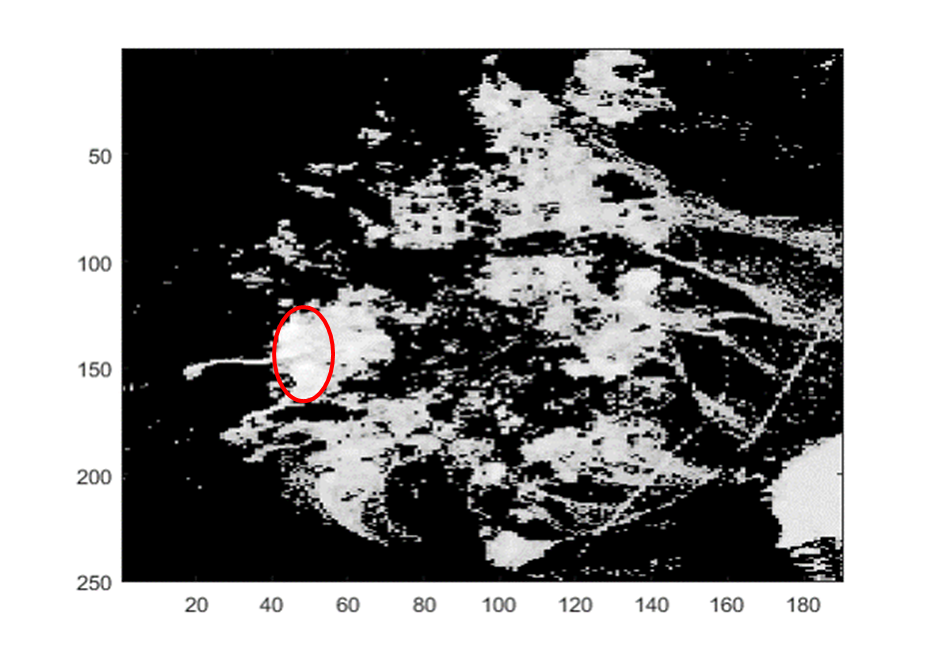}
\endminipage
\minipage{0.25\textwidth}
  \includegraphics[width=\linewidth]{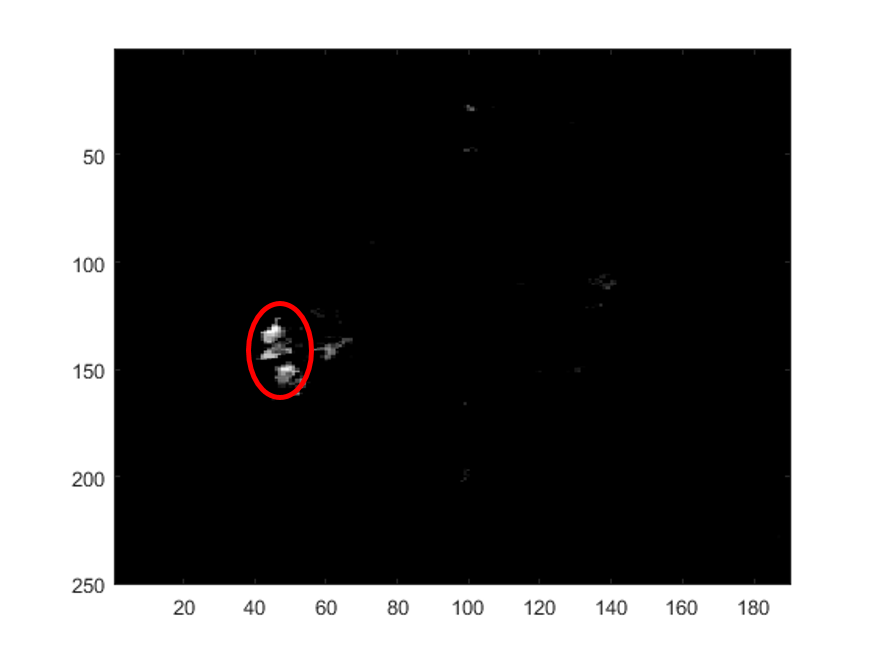}
\endminipage
\caption{Visual separation (we exhibit the mean power in dB over the 186 bands) of the buddingtonite targets using the target dictionary $\mathbf{A}_t$ constructed from the ASTER spectral library. Columns from left to right: original HSI $\mathbf{D}$, low-rank background HSI $\mathbf{L}$, sparse target HSI $\left(\mathbf{A}_t\,\mathbf{C}\right)^T$, and sparse target HSI $\left(\mathbf{A}_t\,\mathbf{C}\right)^T$ after some thresholding.}\label{fig:separation_Buddingtonite}\label{fig:Buddingtonite_separation_ourMethod}
\end{figure*}

\begin{figure}[!htb]
\minipage{0.242\textwidth}
  \includegraphics[width=\linewidth]{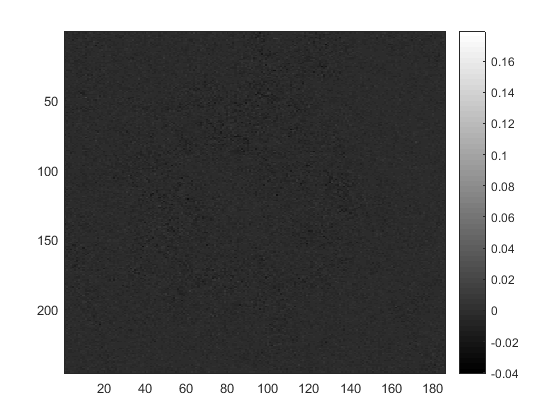}
{\begin{center}%
\vspace{-3mm}
{\bf (a)}
\end{center}}
\endminipage\hfill
\minipage{0.242\textwidth}
  \includegraphics[width=\linewidth]{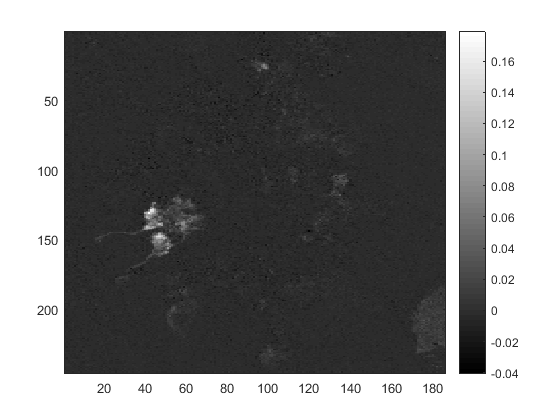}
{\begin{center}%
\vspace{-3mm}
{\bf (b)}
\end{center}}
\endminipage\hfill
\centering
\minipage{0.242\textwidth}
  \includegraphics[width=\linewidth]{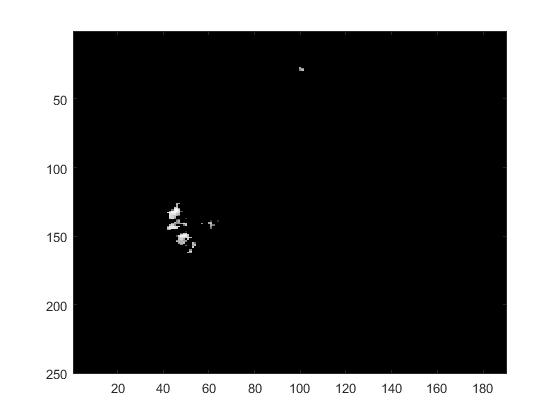}
{\begin{center}%
\vspace{-3mm}
{\bf (c)}
\end{center}}
\endminipage
 \caption{2-D visualization of the buddingtonite target pixels detection. {\bf (a)}: SRBBH detector when $\mathbf{A}_b$ is constructed from $\mathbf{D}$. {\bf (b)} SRBBH detector when $\mathbf{A}_b$ is constructed from $\mathbf{L}$. {\bf (c)} Detection in $\left(\mathbf{A}_t\,\mathbf{C}\right)^T$ for Strategy two (we exhibit the mean power in dB over the 186 bands).}\label{fig:final_results}
\end{figure}
\subsection{Real Experiments}
This section evaluates qualitatively the target detection performances of the SRBBH detector, using a concentric window of size $5\times 5$ on a region of size {\color{red}250 $\times$ 190} pixels taken from the acquired Cuprite HSI. We consider this zone specifically to detect the tectosilicate mineral-type target pixels known as buddingtonite. The mean power in dB over the 186 spectral bands of this zone and the buddingtonite GroundTruth are shown in the fourth row of Fig. \ref{fig:SPCP_example}. 
\\
There are three buddingtonite samples available in the online Advanced Spaceborne Thermal Emission and Reflection (ASTER) spectral library - Version 2.0 \cite{Baldridge09}, and they will form our target dictionary $\mathbf{A}_t$. The ASTER spectral library was released on December 2008 to include data from the USGS spectral library, the Johns Hopkins University spectral library, and the Jet Propulsion Laboratory spectral library.  
\\
Both the HSI and the buddingtonite target samples are normalized to the values between 0 and 1. Fig. \ref{fig:Buddingtonites_samples} shows the buddingtonite target samples taken from the ASTER spectral library.

\subsubsection{Using Strategy One for Detection}
 As a consequence of the decomposition shown in Fig. \ref{fig:Buddingtonite_separation_ourMethod}, the subspace overlap problem shown in Fig. \ref{fig:SPCP_example} (fourth row) is, now, much relieved, as can be seen from Fig. \ref{fig:final_results}. Fig. 10(a) and (b) evaluates qualitatively the SRBBH detection results when $\mathbf{A}_b$ is constructed from $\mathbf{D}$ and $\mathbf{L}$, respectively, using a concentric window of size $5\times 5$. The effectiveness of problem \eqref{eq:convex_model} in improving the target detection is evident.
\subsubsection{Using Strategy Two for Detection}
{Fig. 10(c)} shows the detection of the buddingtonite targets in $\left(\mathbf{A}_t\, \mathbf{C}\right)^T$. The buddingtonite targets are detected with very little false alarms.
\\
As can be seen from all the preceding experiments, both the strategies can deal with the targets of any shapes or targets that occur in close proximity. This is important in many applications, for instance, in the above-stated tectosilicate mineral example, where it is often not possible to fix the window size required in SRBBH.



\section{Conclusion and Future work}
\subsection{Conclusion}
A method based on a modification of RPCA is proposed to separate known targets of interests from the background in hyperspectral imagery. More precisely, we regard the given HSI as being made up of the sum of low-rank background HSI $\mathbf{L}$ and a sparse target HSI $\mathbf{E}$ that should contain the targets of interests. Based on a prelearned target dictionary $\mathbf{A}_t$ constructed from some online spectral libraries, we customize the general RPCA by factorizing the sparse component $\mathbf{E}$ into the product of $\mathbf{A}_t$ and a sparse activation matrix $\mathbf{C}$. This modification was essential to disambiguate the true targets from other small heterogeneous and high contrast regions.

Following the decomposition, the first outlined strategy (Strategy one) addresses the background dictionary contamination problem suffered by the dictionary-based methods such as SRBBH. To do this, the low-rank background HSI $\mathbf{L}$ was exploited to construct $\mathbf{A}_b$. More precisely, for each test pixel in the original HSI, $\mathbf{A}_b$ is constructed from $\mathbf{L}$  using a small concentric window, and all the pixels within the window (except the center pixel) will each contribute to one column in $\mathbf{A}_b$.

An alternative strategy (Strategy two) was to directly use the component $\left(\mathbf{A}_t \mathbf{C}\right)^T$ as a detector. Only the signals that reside in the target subspace specified by $\mathbf{A}_t$ are deposited at the non-zero entries of $\left(\mathbf{A}_t \mathbf{C}\right)^T$. 
Both the strategies are evaluated on both synthetic and real experiments, and the results of which demonstrate their effectiveness for hyperspectral target detection. In particular, they can deal with the targets of any shapes or targets that occur in close proximity and are resilient to most values of target fill-fractions unless they are too small.

\subsection{Some Directions for Future Work}
The paradigm in military applications of hyperspectral imagery seem to center on finding the target but ignoring all the rest.  Sometimes, that rest is important, especially if the target is well matched to the surroundings. As for future enhancements, a likely first step would be to evaluate the proposed modified RPCA model on that challenge in a future study of cuprite. Other promising avenues for further research include the following.
\\
\begin{enumerate}
\item We would like to mention that the selection of $\tau$ and $\lambda$ strongly depends on the HSI used, on the spatial and spectral dimension of the given HSI, on the target of interest to detect, on the location of the target in the image scene, and on the target dictionary $\mathbf{A}_t$. This encourages us to work hard in the future to develop such an automatic selection method for the parameters (i.e., proposing a formula that can take the aforementioned causes as input). 
\item Obviously, we can observe that the $\tau$ to $\lambda$ ratios, as well as the settings of $\tau$ and $\lambda$ for Strategy two, are not similar to those for Strategy one. We highly expect that if one could use directly the $l_{2,0}$ norm (that is, without surrogating it toward the convex $l_{2,1}$ norm), both the detection strategies might have the same parameters settings.
\\
\item Interestingly, what we have not also mentioned before is that the selection of $\tau$ and $\lambda$ depends on the target fill-fraction $\alpha$. During our work, we have done a lot of experiments (omitted here) on the HSI zone used in the synthetic experiments by replacing a fraction $\alpha$ from the background pixel at location $(34, 50)$ by the target $\mathbf{t}$ corresponding to the mean of the six jarosite target samples. We have observed that if one needs to separate the $\alpha\mathbf{t}$ from $(1-\alpha)\mathbf{b}$ using our problem in \eqref{eq:convex_model} (that is, we need that $\alpha\mathbf{t}$ to be deposited in the sparse component and $(1-\alpha)\mathbf{b}$ in the low-rank component), the selection of $\lambda$ will not be unique for all $\alpha$ values. More precisely, the higher the $\alpha$ value is, the more the $\lambda$ value needs to decrease. This is due to the fact that a higher $\alpha$ value implies a more target fraction to separate from the background. However, we highly expect that if one could use directly the $l_{2,0}$ norm (that is, without surrogating it toward the convex $l_{2,1}$ norm), a unique value of $\lambda$ might be chosen for all $\alpha$ values.
\end{enumerate}
In this regard, our future work will mainly focus on the use of other proxies than the $l_{2,1}$ norm (closer to the $l_{2,0}$ norm) which can help to alleviate the $l_{2,1}$ artifact and probably the manual selection problem of $\tau$ and $\lambda$.


\section*{Acknowledgment}
The authors would like to thank Dr. G. A. Swayze from the United States Geological Survey (USGS), for his time in providing them helpful remarks and suggestions, especially about the buddingtonite mineral. 
They would like to thank Dr. Y. Zhang, Dr. B. Du, and Dr. L. Zhang from Wuhan University for providing them the Nuance Cri HSI. 
Finally, they would also like to thank the handling editor and four anonymous reviewers for the careful reading and helpful remarks/suggestions. 

\ifCLASSOPTIONcaptionsoff
  \newpage
\fi

\bibliographystyle{IEEEbib}
\bibliography{biblio_these}

\end{document}